\RequirePackage{amsmath}
\documentclass[]{elsarticle}
\usepackage{float}
\usepackage{graphicx}
\usepackage{footnote}
\usepackage{color}
\usepackage[ruled, linesnumbered, algo2e,  algosection]{algorithm2e}
\usepackage{subcaption} 
\usepackage{caption}
\usepackage{amsfonts}
\usepackage{tikz}
\usepackage{booktabs}
\usepackage{todonotes}
\usepackage{multirow}
\usepackage{verbatim}
\usepackage{breqn}
\usepackage{bm}
\usepackage[toc,page]{appendix}
\usepackage{hyperref}
\usepackage[capitalise]{cleveref}

\hypersetup{colorlinks = true,linkcolor = blue,anchorcolor =red,citecolor = blue,filecolor = red,urlcolor = red,
	pdfauthor=author}
\newcommand{\crossed}{$\times$}
\newcommand{\norm}[1]{\left\lVert#1\right\rVert}

\newcommand{\mgs}{\mathbf{m}^\mathrm{gs}}
\DeclareMathOperator*{\argmax}{arg\,max}

\crefname{appsec}{appendix}{appendices}
\Crefname{appsec}{Appendix}{Appendices}
\begin{document}
\title{Neural collaborative filtering for unsupervised mitral valve segmentation in echocardiography}
\author[1]{Luca Corinzia\corref{cor}}
\ead{luca.corinzia@inf.ethz.ch}
\author[1]{Fabian Laumer}
\ead{fabian.laumer@inf.ethz.ch}
\author[2]{Alessandro Candreva}
\ead{alessandro.candreva@usz.ch}
\author[3]{Maurizio Taramasso}
\ead{maurizio.taramasso@usz.ch}
\author[3]{Francesco Maisano}
\ead{francesco.maisano@usz.ch}
\author[1]{Joachim M. Buhmann}
\ead{jbuhmann@inf.ethz.ch}
\cortext[cor]{Corresponding author}
\address[1]{ETH Zurich, Institute for Machine Learning, Zurich, Switzerland} \address[2]{University Hospital Zurich, Department of Cardiology, Zurich, Switzerland}
\address[3]{University Hospital Zurich, Department of Cardiac Surgery, Zurich, Switzerland}
\begin{abstract}
The segmentation of the mitral valve annulus and leaflets specifies a crucial first step to establish a machine learning pipeline that can support physicians in performing multiple tasks, e.g.\ diagnosis of mitral valve diseases, surgical planning, and intraoperative procedures. Current methods for mitral valve segmentation on 2D echocardiography videos require extensive interaction with annotators and perform poorly on low-quality and noisy videos. We propose an automated and unsupervised method for the mitral valve segmentation based on a low dimensional embedding of the echocardiography videos using neural network collaborative filtering. The method is evaluated in a collection of echocardiography videos of patients with a variety of mitral valve diseases, and additionally on an independent test cohort. It outperforms state-of-the-art \emph{unsupervised} and \emph{supervised} methods on low-quality videos or in the case of sparse annotation.
\end{abstract}
\maketitle 
\section*{Highlights}
\begin{itemize}
    \item Unsupervised segmentation can perform adequately on echocardiography exploiting the low dimensional structure of the video.
    \item Non-linear models, i.e.\ neural collaborative filtering, outperform their linear counterparts by exploiting the high adaptivity of the model.
    \item Our method outperforms supervised methods on low-quality videos and defines a new state-of-the-art method for unsupervised mitral valve segmentation.
\end{itemize}

\section{Introduction}
\label{sec:introduction}
\par The mitral valve (MV) is the largest valve of the heart and safeguards the monodirectional blood flow from the left atrium towards the left ventricle. It is composed of two leaflets, the anterior and the posterior one, that are attached to a fibrous ring known as the mitral annulus.
Its functionality is passively regulated by the pressure gradient between two heart chambers, the left atrium and the left ventricle, with the opening given by the pressure excess of the first compared to the second (during the so-called \emph{diastole} phase), and the closing given in the opposite setting (the \emph{systole} phase).
The diseases affecting the MV apparatus can lead to: (1) a narrowing of the valve orifice, thus impairing the flow across the valve during diastole and provoking the so-called \emph{mitral stenosis}; (2) a defect of the coaptation of the MV leaflets during systole, which causes back-flow into the left atrium known as \emph{mitral regurgitation}. 
This last condition defines the most common cardiac valvular defect and the second most common amenable of surgical intervention \cite{hayek2005mitral}. 
\par Echocardiography (echo) is a medical imaging technique that produces 2D and 3D pictures and videos of the heart using ultrasound waves generated by vibrating piezoelectric crystals, scattered and reflected at the biological tissues interfaces, detected and converted by the machine in digital signals. Echo is the standard imaging tool in the clinical routine to perform the diagnosis of most of the heart diseases and dysfunctions, including MV diseases \cite{hayek2005mitral,zamorano2004real,baumgartner2009echocardiographic}. It is inexpensive, non-invasive and it enables both qualitative and quantitative assessment of the myocardium and the MV functions.
Clinical practice for the assessment of the MV disease requires physicians to manually trace and measure a plethora of diagnostic parameters. The automatic delineation of the MV annulus and of both the MV leaflets (that we will denote in the following by MV segmentation) could enable physicians to create an automated mechanical model of the MV and to improve the quality of visualization and understanding of the MV pathology.
\subsection{Contribution} 
In this paper, we propose NN-MitralSeg, an unsupervised MV segmentation algorithm based on neural collaborative filtering \cite{he2017neural, dziugaite2015neural}, that supports a systematic and fast evaluation of MV health status for medical practitioners. Our method substantially extends our work published in the conference paper \cite{corinzia2019unsupervised} and improves on the Robust Non-negative Matrix Factorization method (RNMF), an unsupervised segmentation method proposed in \cite{dukler2018automatic} with a three-fold contribution: (i) we use a neural collaborative filtering technique \cite{dziugaite2015neural, he2017neural} that generalizes the matrix factorization and accounts for both linear and non-linear contributions of the myocardial wall motion, in combination with a parametrized threshold operator to learn the high dimensional sparse signal that captures the MV; (ii) we leverage the information of both the optical flow of the sparse signal and of the low dimensional time series representation of the echo to delineate the region of interest (ROI); (iii) we apply post-processing algorithms to improve the final MV segmentation. The method outperforms RNMF on a dataset of 39 patients affected with MV dysfunction and mitral regurgitation, and on an additional independent public dataset of 46 patients extracted from the EchoNet-Dynamic dataset \cite{ouyangechonet}. NN-MitralSeg is further compared to the state-of-the-art supervised segmentation method based on the U-Net neural network architecture \cite{ronneberger2015u,costa2019mitral}. U-Net underperforms NN-MitralSeg when trained with up to two annotated frames per videos, while it outperforms it with a higher density of annotation. Nevertheless, its performance on low-quality videos is on par with or worse than NN-MitralSeg at any level of annotation density considered in the study.
\section{Method}
\label{sec:method}
The proposed segmentation model is composed of many stages and follows the structure of other unsupervised methods (see  \cite{corinzia2019unsupervised,dukler2018automatic,zhou2012automatic} and the literature review in \Cref{sec:related_work}). First, the echo video is embedded in a low dimensional space using a factorization technique (e.g.\ non-negative matrix factorization and variations), then the remainder of the factorization, in the following called sparse signal, is used to delineate the ROI and MV segmentation masks. The next sections present every stage of the algorithm in full details.
In this paper, we denote by $a \in \mathbb{R}$, $\mathbf{a} \in \mathbb{R}^d$ and $\mathbf{A} \in \mathbb{R}^{d_1 \times \dots \times d_k}$ respectively a generic scalar, vector, and a $k$-rank tensor with $k \ge 2$.
\subsection{Factorization}
Here, we present the Neural Matrix Factorization model (NeuMF) used as the first stage of the NN-MitralSeg algorithm.
\subsubsection*{Model}
\label{sec:model}
\begin{figure}[ht]
	\centering
	\includegraphics[width=0.85\textwidth]{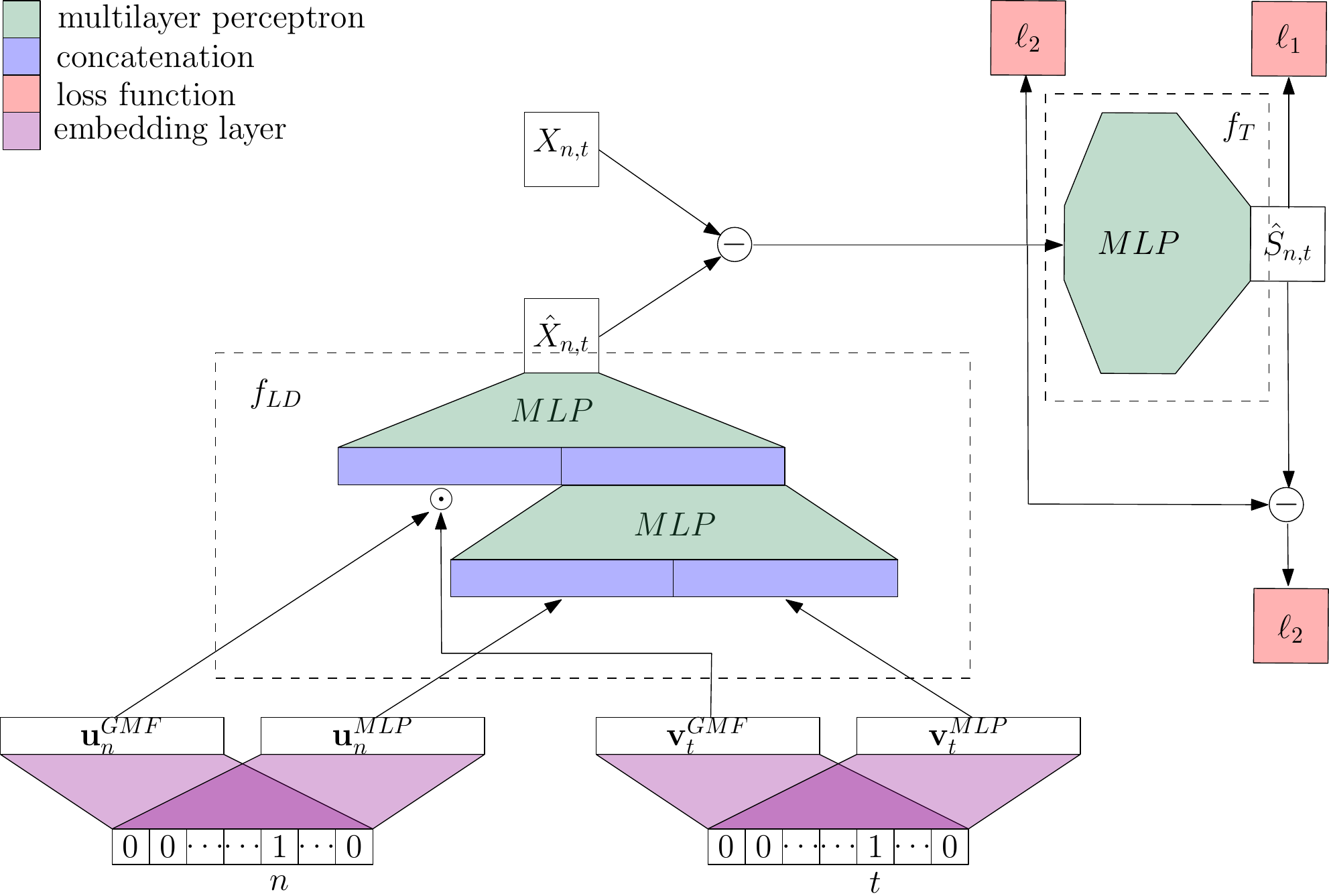}
	\caption{Diagram of the factorization model used in NN-MitralSeg. The network $f_{LD}$ maps the inputs $n$ and $t$ (pixel and frame indices) to the respective latent features $\mathbf{u}_n^{GMF}$, $\mathbf{v}_t^{GMF}$, $\mathbf{u}_n^{MLP}$, $\mathbf{v}_t^{MLP}$ using four different embedding layers (denoted in purple). Such feature vectors are then transformed to the reconstruction $\hat{X}_{n,t}$. The vectors $\mathbf{u}_n^{GMF}$ and $\mathbf{v}_t^{GMF}$ are multiplied element-wise and the resulting vector is concatenated (operation denoted in blue) to the output of a MLP (denoted in green) applied to the vectors $\mathbf{u}_n^{MLP}$ and $\mathbf{v}_t^{MLP}$. Another MLP is then used to produce the approximated signal $\hat{X}_{n,t}$. The threshold operator $f_T$ is then applied to the remainder $\hat{X}_{n,t} - {X_{n,t}}$ to give the sparse signal $\hat{S}_{n,t}$.}
	\label{fig:network_arch}
\end{figure}
Each echo is initially represented as a non-negative tensor $\mathbf{T} \in \mathbb{R}_{+}^{h\times w \times T}$, where $h$ and $w$ are respectively the height and the width of a single frame and $T$ is the number of frames in the video. We reshape each frame of the echo into a column vector and then concatenate all the columns to get a matrix $\mathbf{X}\in \mathbb{R}_{+}^{N \times T}$ where $N = h \cdot w$.
The matrix $\mathbf{X}$ is then embedded in a low dimensional space as follows. We embed each row (pixel) $n \in N$ and each column (frame) $t \in T$ into the low dimensional non-negative latent feature vectors $\mathbf{u}_n, \mathbf{v}_t \in \mathbb{R}_+^{K}$, where $K$ represents the generalized rank of the factorization model. In the case of linear non-negative matrix factorization the element $X_{n,t}$ is approximated with the linear product $\mathbf{u}_n \cdot \mathbf{v}_t$. We generalize the simple linear model parameterizing the interaction between the pixel and the frame feature vectors $\mathbf{u}_n$ and $\mathbf{v}_t$ with a feed forward neural network as 
\begin{equation}
\label{eq:neumf}
\hat{X}_{n,t} = f_{LD}( \mathbf{u}_{n}^{GMF} \odot \mathbf{v}_{t}^{GMF}, \mathbf{u}_{n}^{MLP}, \mathbf{v}_{t}^{MLP}; \bm{\theta}_{LD})
\end{equation} 
where $\odot$ is the element-wise product, $\mathbf{u}_{n}^{GMF}, \mathbf{v}_{t}^{GMF} \in \mathbb{R}^{K}$ are the generalized matrix factorization (GMF) feature vectors and $\mathbf{u}_{n}^{MLP}, \mathbf{v}_{t}^{MLP} \in \mathbb{R}^{K'}$ are additional feature vectors that can embed pixel-wise and frame-wise contributions of the entry $X_{n,t}$ through the multi-layer perceptron (MLP) (see \cite{he2017neural})\footnote{Notice that such model generalizes the $K$-rank non-negative matrix factorization given by
\begin{equation*}
	f_{LD}( \mathbf{u}_{n}^{GMF} \odot \mathbf{v}_{t}^{GMF},\mathbf{u}_{n}^{MLP}, \mathbf{v}_{t}^{MLP}; {\bm{\theta}_{LD}}) = \mathbf{u}_{n}^{GMF} \cdot \mathbf{v}_{t}^{GMF},
\end{equation*} 
hence it can reproduce the latter model with a proper choice of the weights of the MLPs that parametrize the function $f_{LD}$}. 
The function $f_{LD}$ denotes the \emph{low-dimensional} network with weights $\bm{\theta}_{LD}$. The non-negativity of the latent features is imposed using a non-negative activation function. A diagram of the factorization model is given on the top-left part of \Cref{fig:network_arch}. Given the reconstruction $\hat{X}_{n,t}$, the difference between $X_{n,t}$ and $\hat{X}_{n,t}$ serves as the scalar input to the \emph{threshold network} and is transformed to get the scalar output

\begin{equation}
\label{eq:threshold}
\hat{S}_{n,t} = f_T(X_{n,t} - \hat{X}_{n,t}; \bm{\theta}_T)\text{.}
\end{equation}
The threshold network is composed by another MLP with weights $\bm{\theta}_T$ and non-negative activation function (an illustration of the behavior of this function is given in \Cref{fig:thresholds} and well justifies the name given here). \Cref{fig:network_arch}, shows a diagram of the complete architecture.
	
\subsubsection*{Training}
The factorization model is parametrized by the adaptive weights $\bm{\theta}_{LD}$, $\bm{\theta}_{T}$ and the embedding vectors $\mathbf{U} = \{\mathbf{u}_n^{GMF},\mathbf{u}_n^{MLP}\}_{n=1}^{N}$ and $\mathbf{V} = \{\mathbf{v}^{GMF}_t, \mathbf{v}^{MLP}_t\}_{t=1}^{T}$. The low dimensional network and the embedding layers are trained to ensure that the network produces an accurate approximation of $\mathbf{X}$. The objective used for reconstruction reads then
\begin{dmath}
\label{eq:reconstruction_loss}
L_{r}(\bm{\theta}_{LD}, \mathbf{U},\mathbf{V}) = 
\norm{\mathbf{X} - f_{LD}(\mathbf{U},\mathbf{V}; \bm{\theta}_{LD})}_F^2 + \beta\left[ \sum_n
\norm{\mathbf{u}^{GMF}_n}_2^2 + 
\norm{\mathbf{u}^{MLP}_n}_2^2 + 
\sum_{t} 
\norm{\mathbf{v}^{GMF}_t}_2^2 +
\norm{\mathbf{v}^{MLP}_t}_2^2
\right]
\end{dmath}
where $\beta$ denotes a regularization parameter and $\|\cdot\|_F$ is the Frobenius norm. The optimization of \cref{eq:reconstruction_loss} is performed in two consecutive steps, freezing the embedding vectors $\mathbf{U, V}$ while updating $\bm{\theta}_{LD}$, and then freezing the low dimensional network $\bm{\theta}_{LD}$ while updating $\mathbf{U, V}$.
\par The threshold network $f_T$ is applied elementwise on the reconstruction remainder $\mathbf{X} - \mathbf{\hat X}$ to produce the sparse signal $\mathbf{\hat{S}} = f_{T}(\mathbf{X} - \mathbf{\hat X}; \bm{\theta}_T)$.  $f_T$ is trained to suppress the remainder of the reconstruction, using a $\ell_1$ penalization (hence imposing sparsity in the sparse signal and enforcing a threshold-like behaviour in the function $f_T$) while reconstructing the signal as much as possible, keeping fixed both $\bm{\theta}_{LD}$, $\mathbf{U}$ and $\mathbf{V}$. This goal is achieved by optimizing the loss function 
\begin{equation}
\label{eq:sparse_loss}
L_s(\bm{\theta}_T) = \norm{\mathbf{X} - \mathbf{\hat X} - \mathbf{\hat{S}}}_F^2 +
\lambda \norm{\mathbf{\hat{S}}}_1
\end{equation}
where $\lambda$ is the sparsity coefficient and $\|\cdot\|_1$ denotes the $\ell_1$-norm. A summary of the losses used in the model is also given in \Cref{fig:network_arch} by the red boxes, and details on the employed hyperparameters and other details of the optimization routine are specified in \Cref{app:net_specs}. The training dynamics is summarized in \Cref{sec:trainig_dynamics}.

\subsubsection*{Initialization}
The initialization of the model parameters is performed in two distinct way: (i) Random initialization (RI), our first method in subsequent experiments, that relies on Xavier initialization \cite{glorot2010understanding} of all MLPs parameters $\bm{\theta}_{LD}$ and $\bm{\theta}_{T}$ and Gaussian initialization of all the embedding vectors $\mathbf{U}$ and $\mathbf{V}$; (ii) matrix-factorization initialization (MFI), that assigns the output of RNMF \cite{dukler2018automatic} of the echo video $\mathbf{X}$ with rank $K$ to the $K$ dimensional embedding vectors $\mathbf{u}^{GMF}_n$, $\mathbf{v}^{GMF}_t$. 
All other parameters are initialized as in RI. The effects of the two initialization schemes are discussed in \Cref{sec:trainig_dynamics}.

\subsection{Window Detection}
\label{sec:method_window_detection}
After the training of the factorization model, the sparse signal $\mathbf{\hat{S}}$ captures most of the motion expressed by the MV, or by other valves of the heart if they appear in the field-of-view, as well as some speckle noise of the echo; the reader might consult  \Cref{fig:rnmf_failure} for a depiction of the sparse signal for the RNMF method in \cite{dukler2018automatic} and \Cref{fig:training_curves_images} for the sparse signal given by our method. 
\par In line with \cite{dukler2018automatic, zhou2012automatic}, we propose a simple window detection (WD) algorithm based on the computation of the norm of the sparse signal $\hat{\mathbf{S}}$ masked by a sliding window. First, the sparse signal $\hat{\mathbf{S}} \in \mathbb{R}_{+}^{N \times T}$ is reshaped into a 3D array of the same shape as the original video $\mathbb{R}_{+}^{h\times w\times T}$. Denoting by $\{\mathbf{W}_l \in \{0,1\}^{h \times w}\}_l$ the set of all possible rectangular windows of fixed size $M$, we propose a general ROI selection that can be summarized as 
\begin{align}
\begin{split}
l^* = & \argmax_l \sum_{t=1}^T s_t \norm{g(\mathbf{\hat{S}}_t) \odot \mathbf{W}_l}_2^2 \\
 & \text{s.t.}  \norm{\mathbf{W}_l}_0 = M
 \end{split}
\label{eq:window_detection}
\end{align}
where $s_t \in \mathbb{R}_+^T$ is a time-variant weight, $g(\cdot)$ is a generic scalar function applied element-wise on the elements of $\mathbf{\hat{S}}_t$ and $\|\cdot \|_0$ is the zero-norm that counts the non-zero elements. The selection is made between windows spanning the whole 2D frame, with a fixed stride. We recover the WD used in \cite{dukler2018automatic, zhou2012automatic} using time-uniform weights and a simple threshold operator for the function $g$. This WD method is called TO (for threshold operator) in the following. 
\par In this work, we propose two main variations of this general method that leverage also movement information in the choice of the time weights and of the function $g$.
The motion of the MV is much faster compared to the myocardium, even when the myocardium appears in the sparse signal. The norm of the dense optical flow \cite{Farneback:2003} can measure the motion in a video and a large norm is indicative of fast motion. 
Hence we denoted by OF (for optical flow) the WD method obtained using $g(\mathbf{\hat{S}}_t) = V_{opt}(\hat{\mathbf{S}})_t$ in \cref{eq:window_detection}, where $V_{opt}(\cdot)$ is the norm of the dense optical flow. A depiction of the WD algorithms considered so far is given in \Cref{fig:window_detection}.
\begin{figure}[ht]
\centering
\includegraphics[width=1.\textwidth]{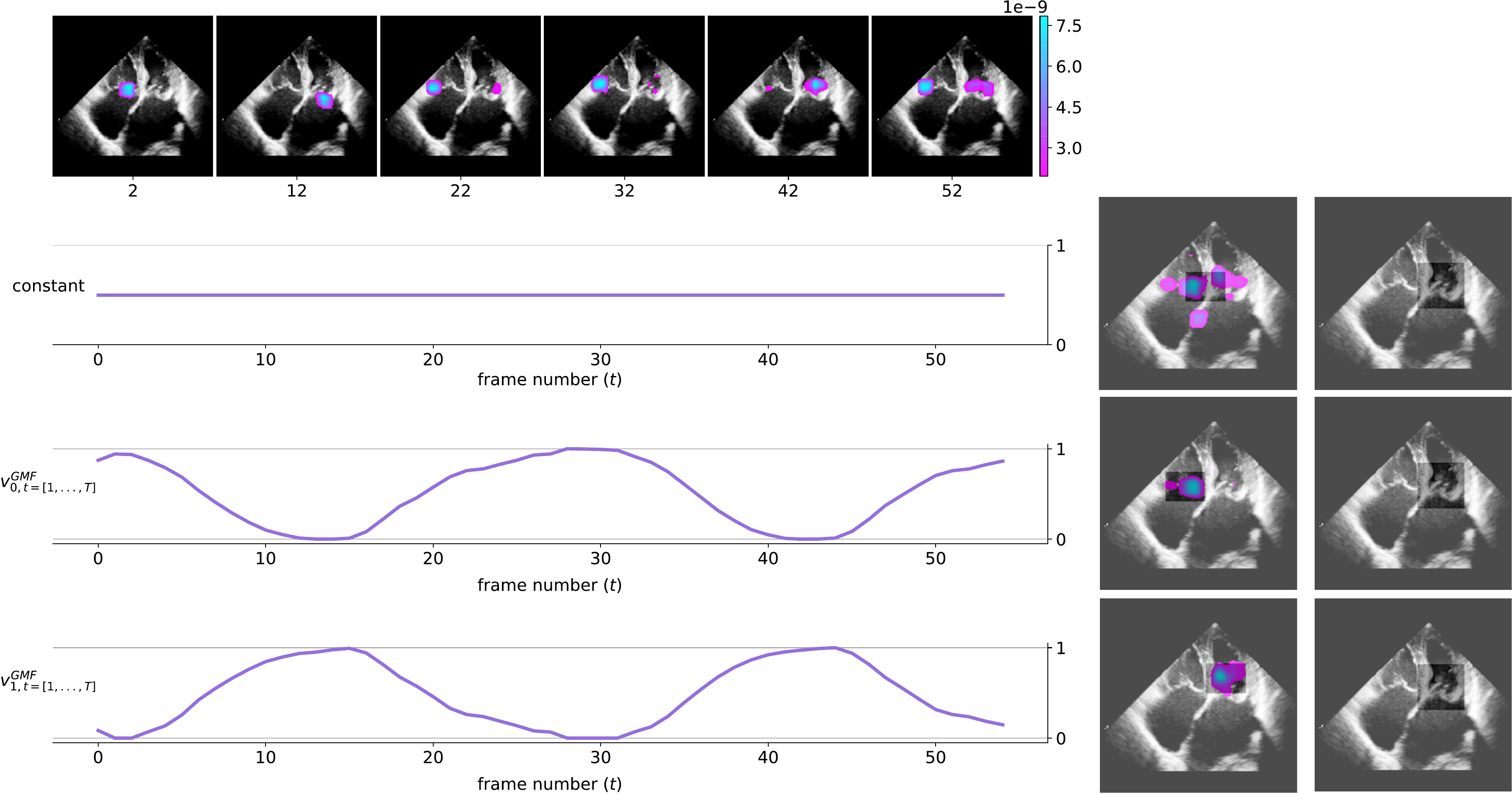}
\caption{Improved window detection algorithm with time-masking. The images at the top depict the dense optical flow of $\mathbf{\hat{S}}$ at different time frames while the three plots indicate different time-variant weighting, namely from top to bottom, uniform weighting and the two components of $\mathbf{v}_{\mathbf{t}}^{GMF}$. In the right two columns, we show the windows selected by the respective time-variant weighting according to \cref{eq:window_detection}, alongside the gold-standard ROI. The uniform (constant) time-variant weighting selects an ROI that is located between the tricuspid and the mitral valve, while the frame embedding components $\mathbf{v}_{0, \mathbf{t}}^{GMF}$ and $\mathbf{v}_{1, \mathbf{t}}^{GMF}$ capture respectively the tricuspid valve and the mitral valve. Best viewed in colours.}
\label{fig:time_masking}
\end{figure}
\par It can be observed that a common source of error for the WD in both TO and OF is the fast movement of other components of the heart, like the tricuspid valve located between the right atrium and right ventricle. Despite being smaller then the MV, this valve can be present in the sparse signal in case of very good quality echos. The window detection used in \cite{dukler2018automatic, zhou2012automatic} and the one described in \cref{eq:window_detection} will in general not be able to clearly distinguish between these two valves. We propose an improved window detection algorithm which identifies the motion of different components of the heart during different states within the heart cycle (e.g.\ the tricuspid and the MV open at different stages, respectively during ventricular diastole and systole). We propose to use the frame embedding vector components $v_{i,t}^{GMF}$ for $i=1,\dots K$ as time-variant weights in \cref{eq:window_detection}. The optimized ROI selection (denoted by time-masking (TM) in the following) then identifies the right-most window from the $K$ windows selected by \cref{eq:window_detection} with the $K$ different time-variant weights (notice that this step can easily be generalized to other prior assumption on the echo video, like echo view, valve to be selected etc.).
In \Cref{fig:time_masking}, the TM method is depicted alongside the behaviour of the frame embeddings $v_{i,t}^{GMF}$ for $K=2$. Notice that the frame embeddings well describe the periodicity of the heart cycle (as noted already in \cite{dukler2018automatic,zhou2012automatic}) and that the WD algorithm either selects the tricuspid valve or the mitral valve as ROI depending which component of $v_{i,t}^{GMF}$ is used as time-variant weights. 
An extensive quantitative analysis of the effect of the different WD methods is given in \Cref{sec:window_detection}.

\subsection{Mitral Valve Segmentation}
\label{sec:mv_segmentation}
The segmentation is consequently performed on the pixels of the sparse signal $\mathbf{\hat{S}}_t$ which are above a fixed threshold and enclosed in the ROI, similarly to \cite{dukler2018automatic, zhou2012automatic} using simple anisotropic 2D diffusion on each frame. We improve this initial segmentation by applying two consecutive post-processing steps: First, we perform erosion and dilation (see \cite{serra2012mathematical} for details on morphological operations) on every frame of the sparse signal. In particular, we apply \textit{opening} (dilation of the erosion) to remove noise from the initial segmentation and confine the segmentation mask to the MV. In a second step, we identify all the connected components of the resulting segmentation mask over the 3D volume and discard the small-size components according to a fixed threshold. A qualitative depiction and quantitative analysis of these post-processing steps are also given in \Cref{sec:post_processing}.

\subsection{Overall algorithm}
The NN-MitrialSeg algorithm that we propose here is given by composing the NeuMF model for factorization introduced in \cref{sec:model}, the WD method based on the optical flow norm (OF) and time-series masking (TM). The MV is then segmented as described in \cref{sec:mv_segmentation}. An extensive evaluation of the performance of the different algorithms is presented in \cref{sec:exp_mv_segmentation}.
\section{Related work}
\label{sec:related_work}
\par MV segmentation in 2D and 3D echo enables automated diagnosis and personalized prognosis of the MV diseases and, therefore, it has received a lot of attention recently. Many early methods are based on active contour algorithms or on other methods that depend extensively on the contribution of human annotators.
Active contour algorithms \cite{isard1996contour,blake2012active}
require medical practitioners to initialize the segmentation algorithm, placing manually a contour close to the desired position in a given frame \cite{mikic1996segmentation} or on multiple frames \cite{shang2008region}. Then the MV is segmented on the given frames optimizing a  predefined energy function, and the mask is propagated over time with the support of the optical flow \cite{mikic1998segmentation} or of a dynamical model of the MV \cite{schneider2011patient}. In \cite{burlina2010patient}, the proposed method leverages both an active contour algorithm that segments the myocardial walls and a thin tissue detector that finds the valve leaflets. In \cite{siefert2013accuracy}, medical practitioners initialize the segmentation denoting multiple points that are then connected using J-splines.
\par The first attempts to design a fully automated MV segmentation algorithm are proposed in \cite{dukler2018automatic,zhou2012automatic}. The 2D echo video is factorized using non-negative 2-rank matrix factorization (NMF) \cite{zhou2012automatic} and its robust extension (RNMF) \cite{dukler2018automatic}. \par The 2-rank factorization captures most of the myocardium wall motion, while the high dimensional sparse signal represents the fast MV movement and the echo speckle noise. Then, the MV is segmented using simple diffusion and thresholding of the sparse signal. Despite producing satisfactory results on high-quality echos, these methods perform below clinically acceptable standards on noise perturbed, low-quality videos, mostly due to the misplacement of the ROI of the MV caused by the low expressiveness of the linear model used.
\par While our segmentation approach is completely unsupervised, supervised methods like those based on neural networks and convolutional architectures \cite{ronneberger2015u, hariharan2015hypercolumns} are rising as the de facto state-of-the-art segmentation method for biomedical 2D imagery matching or exceeding human-level performance. The U-Net architecture \cite{ronneberger2015u} emerged as one of the best performing architecture for medical image segmentation and enables clinicians to learn models with very good generalization performance from only a few annotated samples. It has been been applied in the context of MV segmentation in \cite{costa2019mitral} and it is here reviewed and tested for comparison.
\section{Experiments and results}
\label{sec:results}
\subsection{Dataset description}
\label{sec:data_set}
A total of 39 transthoracic echos were obtained from the MitraSwiss Registry, a Swiss-wide prospective registry which includes patients undergoing percutaneous mitral valve repair using the MitraClip system. All patients had moderate-to-severe (3+) or severe (4+) mitral regurgitation of functional or degenerative origin as graded according to current recommendations of the American Society of Echocardiography \cite{zoghbi2003recommendations}. Imaging data were processed in an anonymized way and all patients provided written informed consent to be included in the study cohort.
Only 4-chamber echo views are used, and for every echo, a rectangular window around the MV (ROI) and three selected frames were annotated by an expert medical doctor. All echos have different $w$, $h$ and $T$ dimensions ranging $w=600-1007$, $h=579-732$, $T=39-159$. The frame rates vary from 25 Hz up to 60 Hz with an average rate of 43 Hz. The height is first zero-padded to match the size of the width and then the spatial dimensions are down-sampled to a fixed size of $400\times400$ pixels.
Additionally, we evaluated the proposed method on another 46 echocardiographic videos from the publicly available EchoNet-Dynamic dataset (see \cite{ouyangechonet} for further details) of the characteristically lower resolution of only 112x112 pixels. The performance on this dataset is reported in \Cref{tab:echonet_results} in the \Cref{app:echonet}.
\subsection{Segmentation metrics} 
In the following, we denote by $\mathbf{m}$ and $\mgs \in \{0,1\}^{w \times h}$ two binary masks, with the gold-standard being denoted by $\mgs$. The window detection accuracy $I$ is defined as the percentage of pixels in the computed ROI that intersect the gold standard mask, formally $I(\mathbf{m}, \mgs) = \frac{|\mathbf{m} \cap \mgs|}{|\mathbf{m}|}$. Note that in this specific task, the window sizes are fixed and not inferred by the model, hence the accuracy is a reliable measure of performance. Other standard metrics for segmentation tasks are the intersection over Union, that reads instead $IoU = \frac{|\mathbf{m} \cap \mgs|}{|\mathbf{m} \cup \mgs|}$, and the S\o rensen Dice reads, $DC = 2\frac{|\mathbf{m} \cap \mgs|}{|\mathbf{m}| + |\mgs|}$.
\subsection{Limitations of RNMF}
\begin{figure}[h]
\centering
\includegraphics[width=0.8\textwidth]{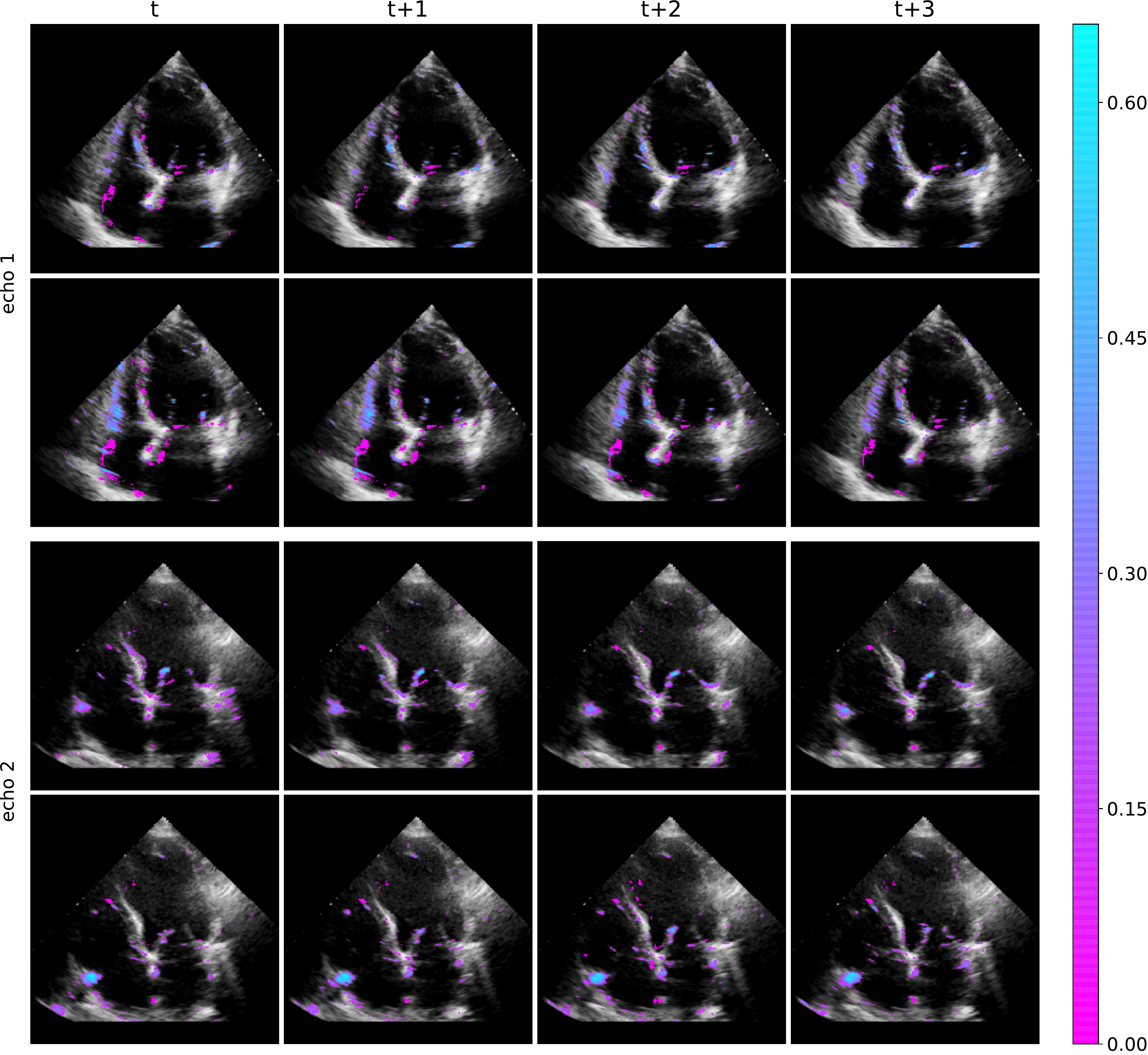}
\caption{Nonlinear motion of the myocardium captured by the sparse signal $\mathbf{\hat{S}}$ in the RNMF method \cite{dukler2018automatic} in two echos. For every echo, four successive frames are reported in the valve closing phase (top row) and opening phase (bottom row). The sparse signal captures the mitral valve movement as well as considerable portions of the myocardium (namely the right ventricle movement for the echo on the top, right atrial movement for the echo at the bottom) in both the opening and closing of the MV. Best viewed in colours.}
\label{fig:rnmf_failure}
\end{figure}
In \Cref{fig:rnmf_failure}, we show two cases of ROI misplacement for the RNMF method \cite{dukler2018automatic}. ROI misplacement in the RNMF technique is caused by limitations of the rank-2 matrix factorization, i.e., some of the myocardium movement cannot be separated by a linear decomposition. The mixture of the end-systole and diastole fail to capture some nonlinear movement of the myocardium and this portion of the movement is encoded in the sparse matrix, which essentially corrupts the window detection algorithm. An example of this myocardium movement can be seen in two different echos and two different phases of the cardiac cycle, namely opening and closing of the MV.
\subsection{NeuMF training dynamics}
\label{sec:trainig_dynamics}
\begin{figure}[h]
\centering
\includegraphics[width=0.7\textwidth]{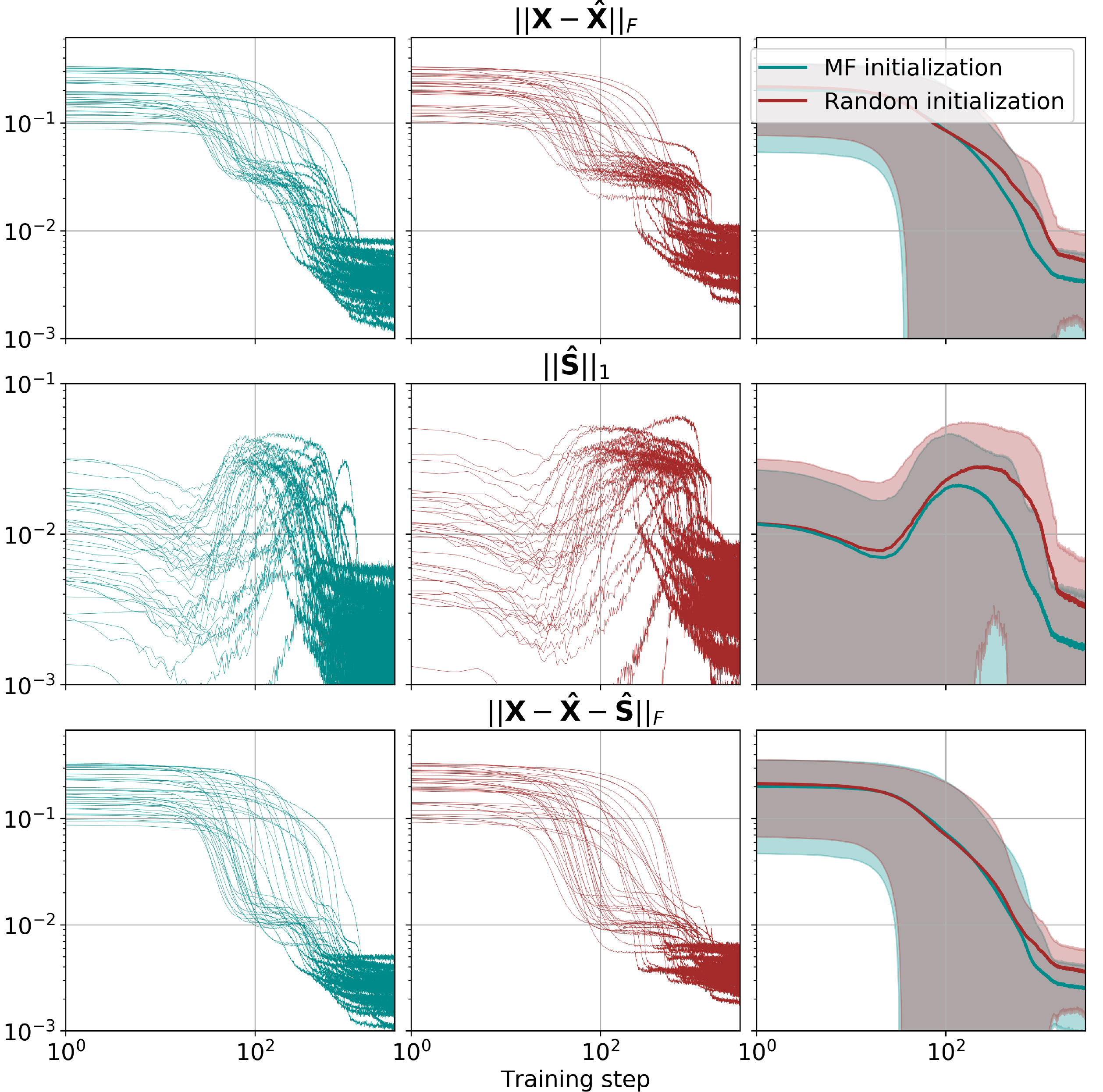}
\caption{Value of the losses (normalized over the batch size) for all the echos during training. From top to bottom we report respectively the reconstruction loss, the $\ell_1$ norm of the sparse signal, and the norm of the reminder when also the sparse signal is considered. From left to right we show the curves for MFI, RI and the average over all the echos. In this plot and in the followings, shaded areas encompass a standard deviation interval around the mean. Log scale is used on both axes.}
\label{fig:training_curves}
\end{figure}

\begin{figure}[h]
\centering
\begin{minipage}{0.7\textwidth}
\begin{subfigure}[t]{0.34\textwidth}
\includegraphics[width=\textwidth]{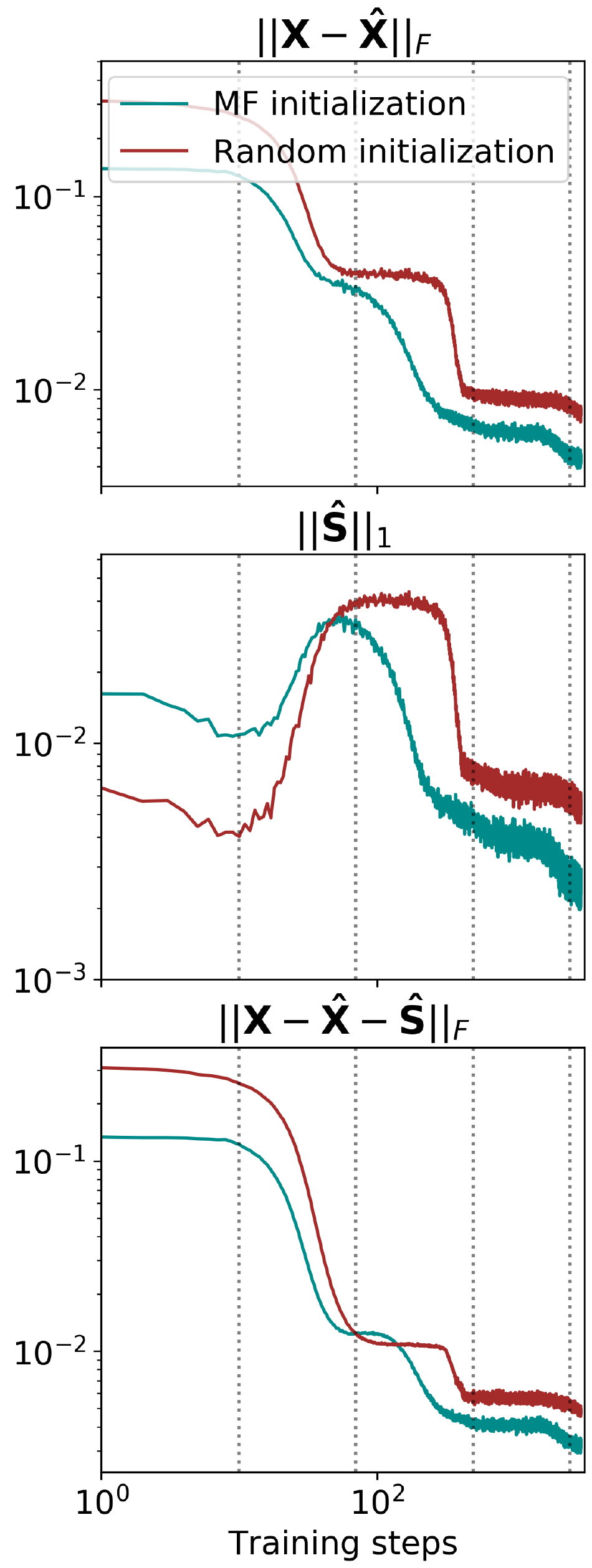}
\caption{Training curves for an individual echo.}
\label{fig:training_curves_single}
\end{subfigure}
\quad
\begin{subfigure}[t]{0.65\textwidth}
\includegraphics[width=\textwidth]{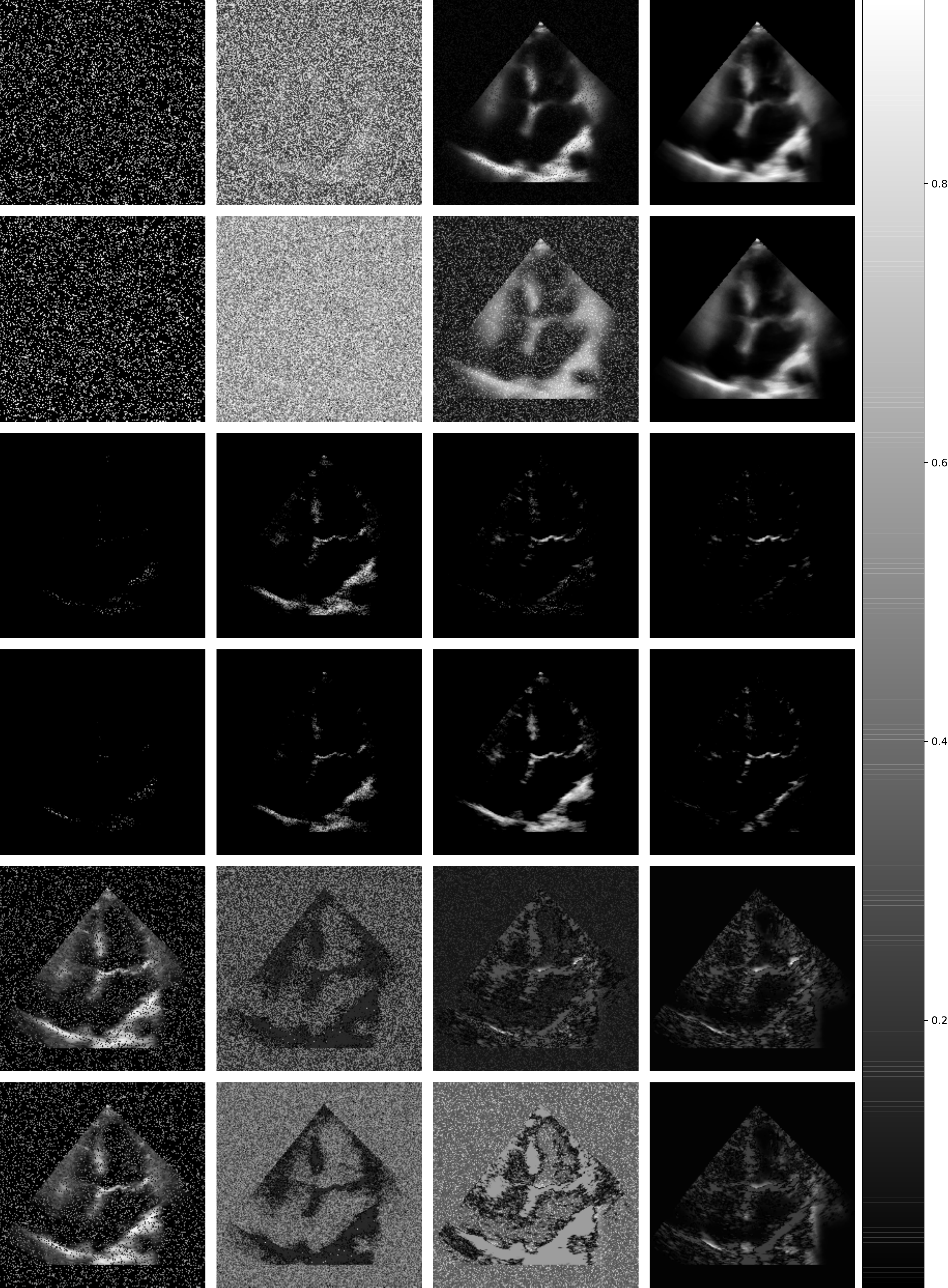}
\caption{Reconstruction ($\hat{\mathbf{X}}$), sparse signal ($\hat{\mathbf{S}}$) and full remainder ($\mathbf{X} - \mathbf{\hat{X}} - \mathbf{\hat{S}}$) at four stages of training. For visualization purposes, only one frame of the video is depicted.}
\label{fig:training_curves_images}
\end{subfigure}
\end{minipage}
\caption{(a) The same quantities as in \Cref{fig:training_curves} are reported for a single echo showing the characteristic four phases of training indicated by the vertical dashed lines, alongside (b) the depiction of the myocardium, sparse signal, and remainder at the four training steps indicated by the vertical lines. For each plot in (a), the top row corresponds to the MF initialization and the bottom row to the random initialization. All the plots are in log-log scale. Full details in the main text. Best viewed in colours.}
\label{fig:training_curves_all}
\end{figure}

The NeuMF method presented in \Cref{sec:model} shows an interesting 
dynamics during training, as can be seen in \Cref{fig:training_curves}. There, we report the value of the losses $\ell_{2x} = \norm{\mathbf{X} - \mathbf{\hat{X}}}_F$, $\ell_1 = \norm{\mathbf{\hat{S}}}_1$ and $\ell_{2xs} = \norm{\mathbf{X} - \mathbf{\hat{X}} - \mathbf{\hat{S}}}_F$ over training for all the echos considered. In \Cref{fig:training_curves_all}, the neural network training dynamics for a single echo is depicted. Four different phases are distinguishable. At the beginning of the training (first dashed vertical line), all quantities are decreasing at a low rate, with the reconstruction $\mathbf{\hat{X}}$ being mostly noise for both RI and MFI as it can be seen in \Cref{fig:training_curves_images}. Then, we observe a second phase with a steep decrease of both the reconstruction losses $\ell_{2x}$ and $\ell_{2xs}$ and the increase of the $\ell_1$ loss, that is also observed in the images from the second column of \Cref{fig:training_curves_images}, with the reconstruction and the sparse signal showing spatial structure. After a plateau of all the quantities, both the reconstruction losses and the sparse signal $\ell_1$ loss decrease. From the third column of \Cref{fig:training_curves_images} we can observe that the sparse signal is focusing on the MV region, with the MFI being sparser and with a smaller contribution from the myocardium movement then the RI. This effect can be quantitatively assessed in \Cref{fig:training_curves} where we show the average losses. The MFI reaches better reconstruction in both $\ell_{2x}$ and $\ell_{2xs}$ for every echo considered.
~
\par
\begin{figure}[h]
	\centering
	\begin{minipage}{0.8\textwidth}
	\begin{subfigure}{1\textwidth}
		\includegraphics[width=\textwidth]{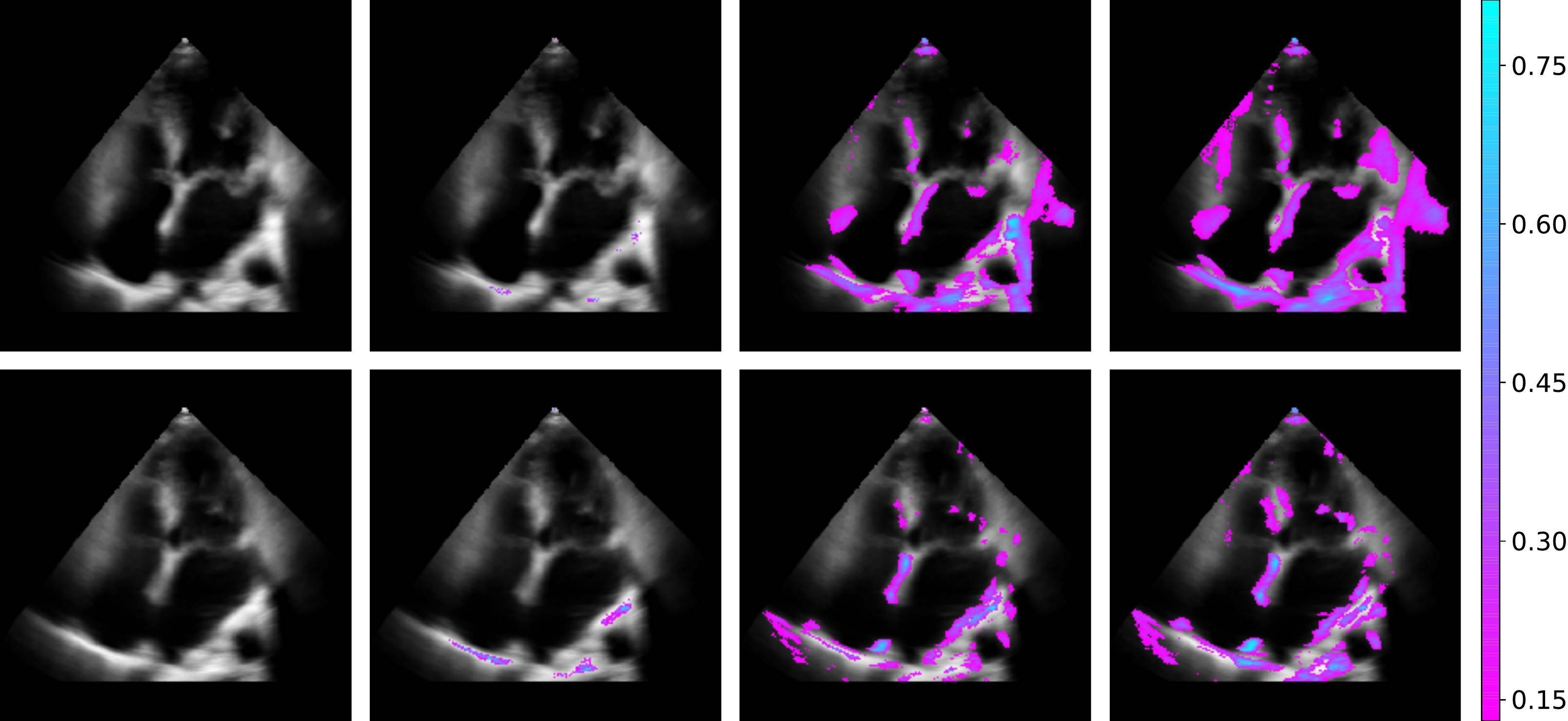}
		\caption{Depiction of the $\mathbf{u}^{GMF}_n$ change during training with MFI}
		\label{fig:gmfu_mfi}
	\end{subfigure}
	\begin{subfigure}{1\textwidth}
		\includegraphics[width=\textwidth]{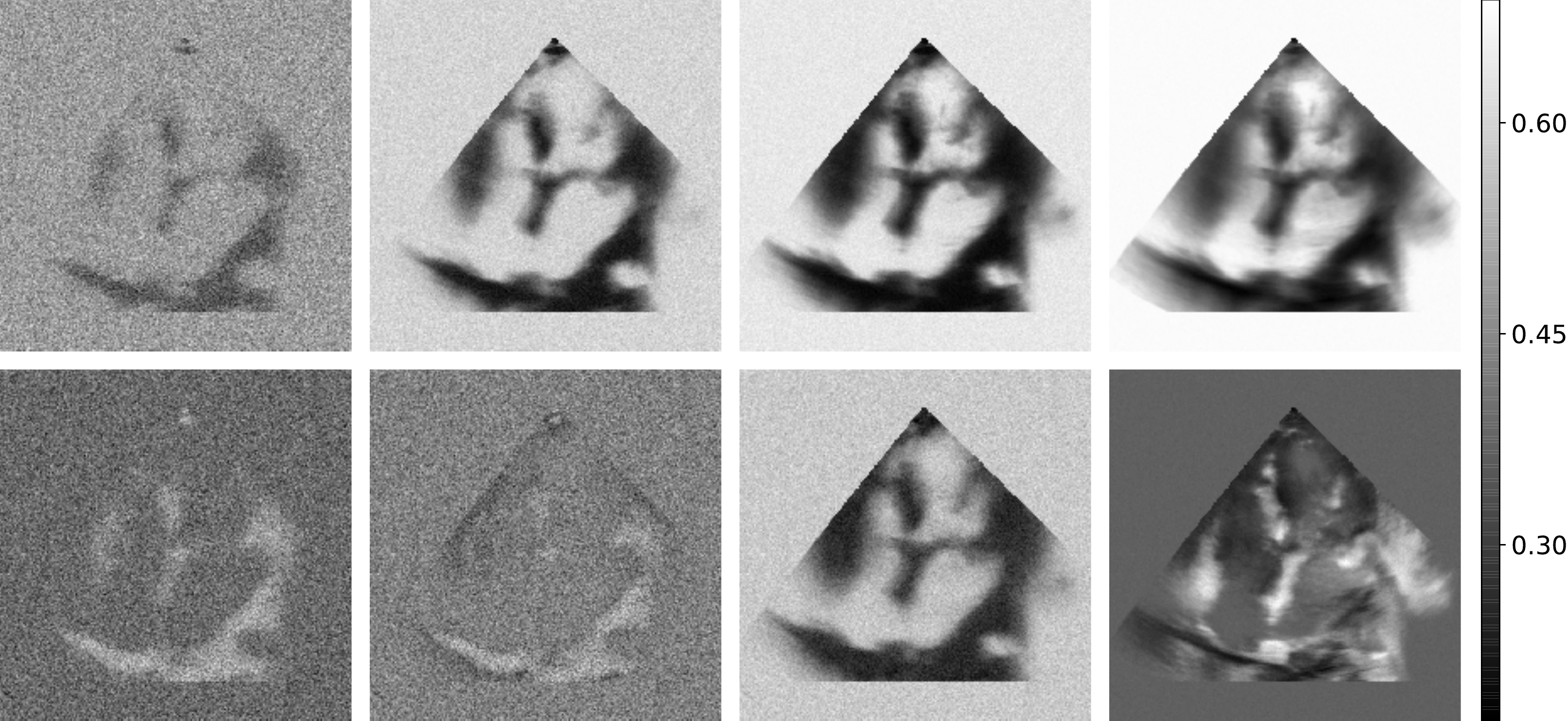}
		\caption{Depiction of $\mathbf{u}^{GMF}_n$ during training with RI}
		\label{fig:gmfu_ri}
	\end{subfigure}
	\end{minipage}
	\caption{Embedding vectors $\mathbf{u}^{GMF}_n$ during training, at the four steps indicated by the dashed vertical lines in \Cref{fig:training_curves_all}. The embedding vectors $\mathbf{u}^{GMF}_n \in \mathbb{R}_+^K$, with $n=1,\dots,w \cdot h$ are reshaped into a $K\times w\times h$, and reported as $K=2$ slices of shape $w\times h$ as images. In (a) we report the embedding vector changes during training, superimposed to the embedding at initialization for the MFI. In (b) we report the same vectors for the RI. Best viewed in colours.}
	\label{fig:gmf_u}
\end{figure}

 As a further analysis of the training dynamics, we monitor the embedding vectors $\mathbf{u}^{GMF}_n$ for both initialization schemes in \Cref{fig:gmfu_mfi} and \Cref{fig:gmfu_ri}. Two main points are noteworthy to observe: (i) For both schemes, the two rows that account for the two dimensions ($K=2$) of the vectors $\mathbf{u}^{GMF}_n$ assume different appearances and have both a spatial structure, despite no spatial constraint or information is explicitly embedded in the model (the input is unrolled into a 2D array with spatial structure; a model with explicit spatio-temporal structure encoded by an additional loss is analyzed in \Cref{app:gaussian_smoothing} and its performance are reported in \Cref{tab:original_cohort_results}); hence, the dimension of the vector is exploited to explain the variability of the data.
(ii) In the case of MFI in \Cref{fig:gmfu_mfi}, the change of the embedding vector, visualized with colours, focuses on the borders of the myocardium walls at the end of the training.  This behaviour shows that the complexity of the factorization model manages to express the non-linear motion of the myocardium that is not captured by the linear model. 
Further embeddings (namely $\mathbf{v}^{GMF}_n$, $\mathbf{u}^{MLP}_n$) are reported in the \Cref{app:embeddings} for completeness.
\subsection{Window Detection performance}
\label{sec:window_detection}
\begin{figure}[]
	\centering
	\begin{minipage}{0.8\textwidth}
		\hspace{0.14cm}
		\includegraphics[width=0.865\textwidth]{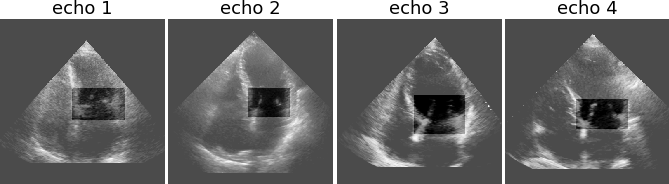}
		\par \medskip
		\includegraphics[width=0.995\textwidth]{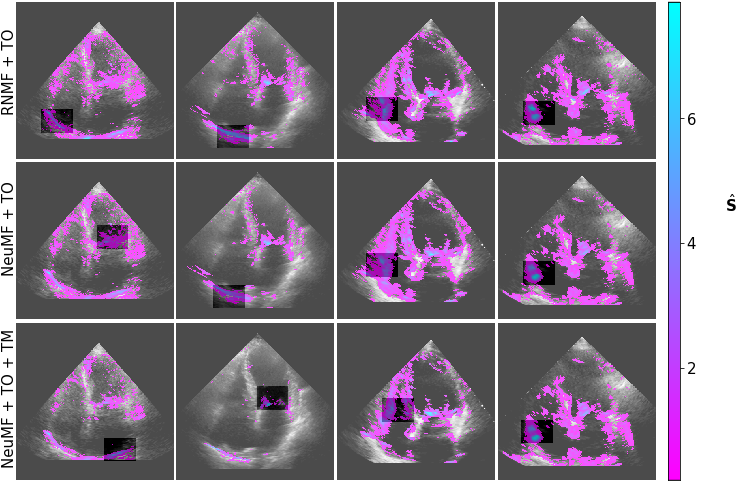}
		\includegraphics[width=\textwidth]{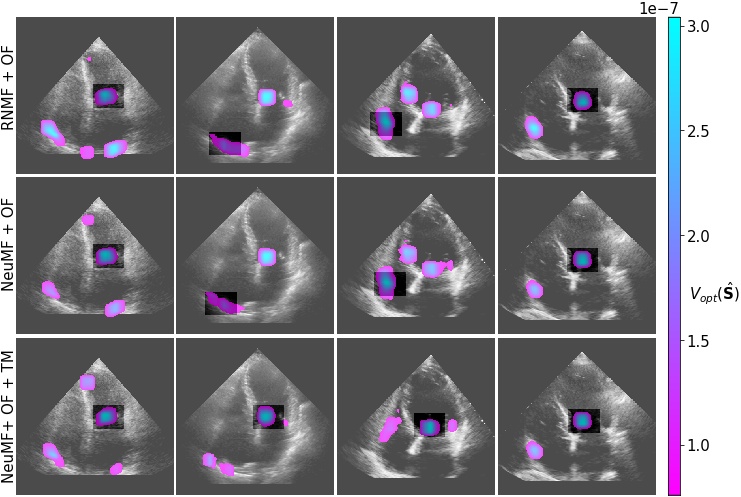}
	\end{minipage}
    \caption{Window detection algorithm outputs on four different echos. All the ROIs are indicated as the shaded area in each frame. The top row corresponds to the gold-standard, and every column represents one single echo. Rows from two to four represent respectively the algorithms RNMF + TO, NeuMF + TO, NeuMF + TO + TM, rows from five to seven represent respectively the algorithms RNMF + OF, NeuMF + OF, NeuMF+ OF + TM. Best viewed in colours. Full details in the main text.}
	\label{fig:window_detection}
\end{figure}

The performance of the different WD methods considered in \Cref{sec:method_window_detection} crucially influences the success of mitral valve segmentation. We first observe in \Cref{fig:window_detection} the qualitative behaviour of the RNMF method in four failure cases (rows two and five). We can see that in all cases the failure is caused by strong myocardium movement that is not captured by the linear model, and hence is present in the sparse signal. The NeuMF method alone (row three) decreases the amount of signal involved in the myocardium movement using a higher capacity factorization model, and this leads to mitigate some of the failure cases of RNMF (see  \Cref{tab:original_cohort_results} for an extensive quantitative evaluation). The WD methods based on the optical flow (OF) are presented in rows five to seven and show a sharper signal of moving pixel in the sparse signal. We can see that this mitigates most of the RNMF failures, giving always a stronger signal in the MV area, alongside however other areas like those capturing the tricuspid valve and myocardium movements. These different contributions are separated by the TM method that is reported at the bottom row, leading to a satisfactory ROI delineation in these four echos considered.
\par The quantitative assessment is performed in different stages. In \Cref{tab:original_cohort_results}, we report the number of WD success cases ($I_{65}$ and $I_{65}$) for two accuracy thresholds (namely 0.65 and 0.85 that represent respectively a level of satisfactory and good WD) and the average $IoU$ and accuracy $I$ for the different methods distinguished by the factorization method (RNMF, NeuMF with RI and MFI), the type of window detection function used (threshold operator TO or optical flow OF) and whether TM has been applied. We observe that NN-MitralSeg outperforms all other methods, which includes the state-of-the-art RNMF as described in \cite{dukler2018automatic}, in all the metrics considered, except for $I_{85}$. Interestingly, applying either NeuMF or the optical flow WD individually in the MV segmentation framework does not increase the WD performance relative to RNMF. The low expressiveness of the linear model in RNMF leaves a strong presence of the myocardium movement in the sparse signal. The dense optical flow is then computed on a sparse signal with a mild amount of myocardium movement and hence its performance decreases as it can be seen in row one and three of \Cref{tab:original_cohort_results} and in \Cref{fig:window_detection}. On the contrary, NeuMF has a high expressiveness and captures most of the myocardium movement in the echo. However, the high expressiveness also captures a small amount of the MV movement, hence the WD method based on the original sparse signal has a lower performance in NeuMF compared to RNMF. 
The combination in NN-MitralSeg uses the benefits of NeuMF, optical flow and TM to achieve better performance. The high expressiveness of NeuMF reduces the brightness of the pixels in the MV region in $\hat{\mathbf{S}}$ but captures a large portion of the myocardium movement. The dense optical flow calculation on $\hat{\mathbf{S}}$ is not corrupted by the myocardium movement, resulting in the best performance in window detection among all methods.
The random initialization method is also reported in \Cref{tab:original_cohort_results} and shows a substantial gap in performance compared to the MFI method.

\begin{table}[]
\caption{Summary of results of both window detection (WD) and mitral valve (MV) segmentation methods. The second and third columns refer to the WD method, and indicate whether the WD uses thresholding applied to the sparse signal (TO) or its dense optical flow (OF), and whether time masking (TM) is performed. The four centre columns report the performance of the respective method in the WD task, namely from left to right the number of samples with accuracy ($I$) larger than 0.65 and 0.85 ($I_{65}$, $I_{85}$), average Intersection over Union ($IoU$) and average accuracy. The two rightmost columns report the average performance of the valve segmentation task, according to $IoU$ and Dice score $DC$. The number in parenthesis is the performance obtained with a post-process of the segmentation labels according to \Cref{sec:mv_segmentation}. The method introduced in \cite{dukler2018automatic} corresponds to the first row (RNMF + TO). NN-MitralSeg corresponds to the last two rows dedicated to the unsupervised methods (NeuMF + OF + TM) with the Gaussian smoothing variation (GS) introduced in \Cref{app:gaussian_smoothing}. We report the U-Net performance dependent on the number of labelled frames per echo available for the training ($l$). The performance on the WD task is not dependent on the number of labels, hence is reported only once. The last two rows describe the performance of the active contour method (see \Cref{app:active_contour} for details) dependent on the initialization method (init.) used (given by the naive ROI or its improvement acc.). Since the algorithm requires a manually initialized bounding box we omit reporting the window detection performance.}
\centering
\small
\setlength\tabcolsep{4pt}
\begin{tabular}{ccc | c c c c | c c c c}
\toprule
\textbf{Method}&\bfseries WD& \bfseries TM & $\bm{I_{65}}$ & $\bm{I_{85}}$ & $\bm{IoU}$ & $\bm{I}$ & $\bm{IoU}$ & $\bm{DC}$ \\
\midrule
\multirow{4}{*}{\textbf{RNMF}} & TO & \crossed & 35  & 28  & 0.387  & 0.822  & 0.258 (0.263) & 0.390 (0.395)\\
& TO & \checkmark & 35 & 32 & 0.422 & 0.871 &  0.270 (0.277) & 0.406 (0.413) \\
& OF & \crossed & 33  & 27  & 0.383 & 0.811  & 0.250 (0.256) & 0.382 (0.387) \\
& OF & \checkmark & 36 & \bfseries 35  & 0.426 & 0.886 &  0.283 (0.293) & 0.424 (0.434)\\
\\
\multirow{4}{*}{\shortstack{\textbf{NeuMF} \\ \textbf{RI} }}& TO & \crossed & 24 & 13 & 0.275 & 0.612 &  0.197 (0.201) & 0.300 (0.305)\\
& TO & \checkmark & 25 & 20 & 0.301 & 0.639 &  0.206 (0.210) & 0.310 (0.316)\\
& OF & \crossed & 31 & 21 & 0.349 & 0.745 &  0.229 (0.230) & 0.347 (0.346) \\ 
& OF & \checkmark & 26 & 20 & 0.331 & 0.694 & 0.210 (0.218) & 0.321 (0.330)\\
\\
\multirow{4}{*}{\shortstack{\textbf{NeuMF} \\ \textbf{MFI}}} & TO & \crossed & 34  & 30  & 0.384  & 0.816  & 0.300 (0.310) & 0.440 (0.449)\\
& TO & \checkmark & 36 & 32 & 0.449 & 0.929 &  0.304 (0.317) & 0.451 (0.462)\\
& OF & \crossed & 34 & 24 & 0.368 & 0.787 & 0.262 (0.271) & 0.393 (0.402)\\ 
& OF & \checkmark & \bfseries 39 & 33 & \bfseries 0.453 & \bfseries 0.939 &  \bfseries 0.326 (0.339) & \bfseries 0.482 (0.495)\\
\shortstack{\textbf{NeuMF}\\\textbf{MFI GS}} & OF & \checkmark & 35 & 30 &  0.403 &  0.863 &  0.306 (0.314) &  0.447 (0.461) \\
\midrule
& & $\bm{l}$   & & & & & & \\
\midrule
\multirow{3}{*}{\textbf{U-Net}}  &  & 1 & \multirow{3}{*}{38} & \multirow{3}{*}{36} & \multirow{3}{*}{0.468} & \multirow{3}{*}{0.967} & 0.162 (0.154) & 0.237 (0.222) \\
& & 2 & & & & & 0.300 (0.274) & 0.415 (0.382) \\
& & 3 & & & & & 0.471 (0.458) & 0.615 (0.599)  \\
\midrule
& & \bfseries init & & & & & & \\
\midrule
\multirow{2}{*}{\textbf{AC}}  &  & ROI & \multirow{2}{*}{-} & \multirow{2}{*}{-} & \multirow{2}{*}{-} & \multirow{2}{*}{-} & 0.152 (-) & 0.259 (-) \\
& & acc & & & & & 0.311 (-) & 0.459 (-) \\
\bottomrule
\end{tabular}
\label{tab:original_cohort_results}
\end{table}

\begin{figure}[ht]
	\centering
	\includegraphics[width=\textwidth]{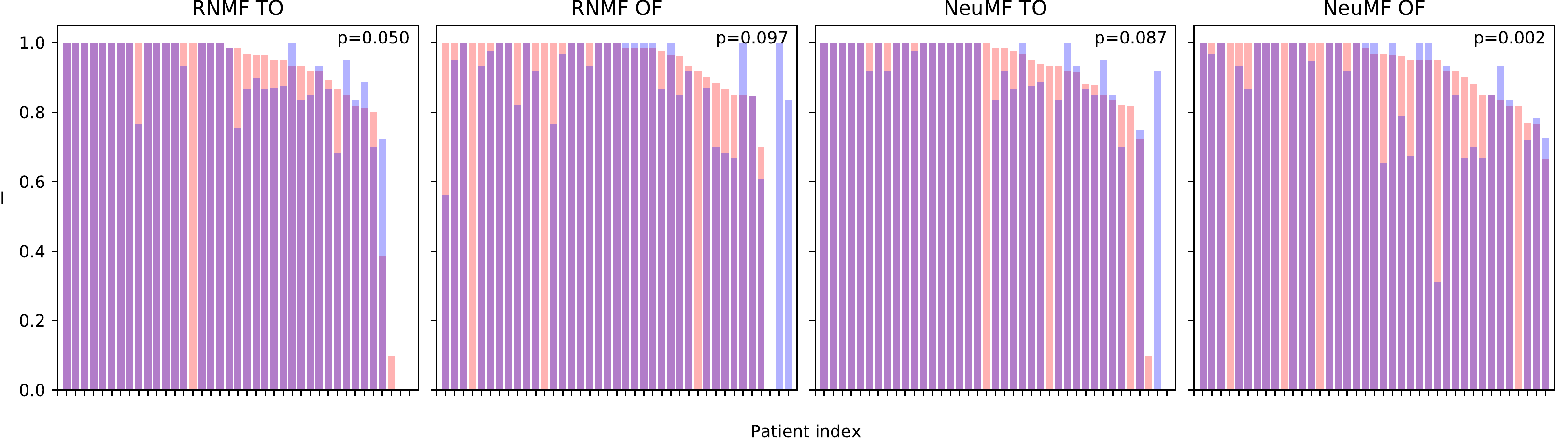}
	\caption{Effect of the time masking method on the performance of different WD methods for each video. The title of each plot reports the factorization method and the processing of the sparse signal used (TO or OF) on which the TM is applied. Red and blue bars represent respectively the method with and without TM. Accuracy $I$ is sorted according to the method that uses TM, and on the top right of each plot, the p-value for a one-sided t-test is reported. Better viewed in colours.}
	\label{fig:effect_time_masking}
\end{figure}

The effect of the TM method is reported in \Cref{fig:effect_time_masking}. The TM method increases the performance of the WD method with a statistically significant margin and notably uniformly over all the echos (it increases the performance on more than 90\% of the echos in all four WD method considered). 
\par For completeness, we also report in \Cref{tab:original_cohort_results} the performance of two other segmentation methods, namely, the state-of-the-art supervised deep learning method based on the U-Net architecture \cite{ronneberger2015u,costa2019mitral}, and an active contour method \cite{kass1988snakes} which requires manual initialization. The U-Net, despite being a supervised method and hence not comparable with the other methods considered so far, provides a strong benchmark for the task. It can be seen that on average it outperforms the NN-MitralSeg. It is however less robust and highly influenced by the echo quality, underperforming on the $I_{65}$ score and low-quality video as it can be seen in \Cref{fig:nn_mitralseg_vs_literature} (two plots on the right). Full details on the implementation of the U-Net method is given in \Cref{app:unet_specs}. More details about the active contour method are presented \Cref{app:active_contour}.

\subsection{Mitral Valve segmentation performance}
\label{sec:exp_mv_segmentation}
\begin{figure}
	\centering
	\includegraphics[width=\textwidth]{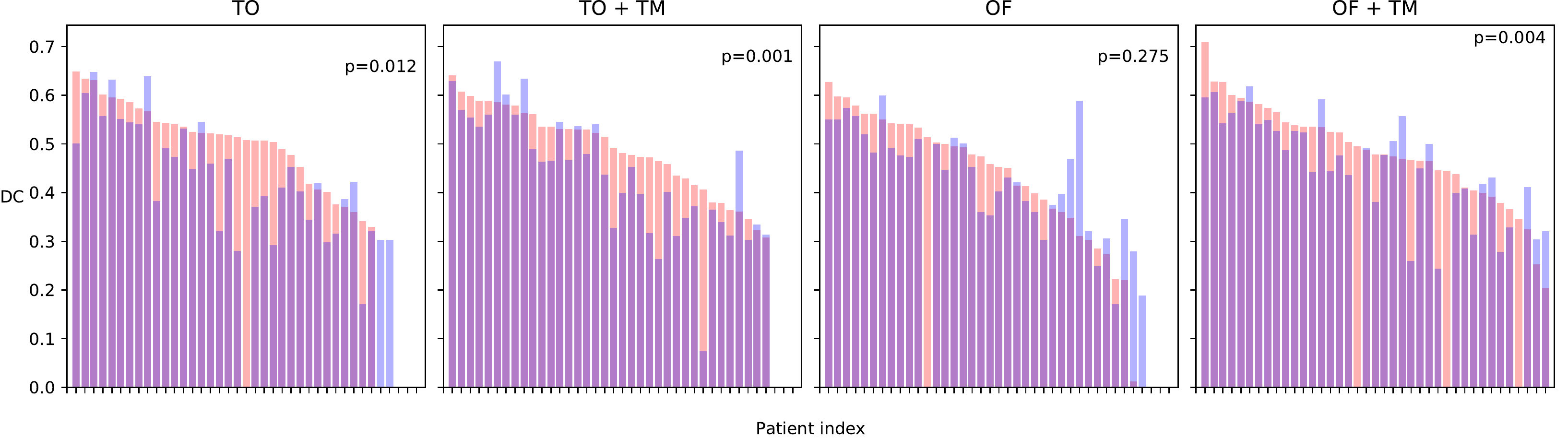}
	\caption{Effect of the NeuMF on the performance of different segmentation methods reported as in \Cref{fig:effect_time_masking}. Red and blue bars are respectively NeuMF and RNMF based methods. The $DC$ score is sorted according to the NeuMF factorization. On the top right of each plot, the p-value for a one-sided t-test is reported.}
	\label{fig:effect_neumf}
\end{figure}

The MV segmentation performance is assessed here using IoU and the dice coefficient $DC$ with respect to the gold standard available as introduced in \Cref{sec:data_set}. In \Cref{tab:original_cohort_results}, we report both scores for all the methods considered here. We can observe the same trends as for the WD task, with a distinction being the effect of the NeuMF, that increases the performance in the MV segmentation by a statistically significant margin. For some videos, the RNMF manages to place the ROI more accurately than the WD based on NeuMF factorization. Even for these videos, the segmentation performance of NeuMF exceeds the RNMF method. This effect is documented in \Cref{fig:effect_neumf} and it is related to the sparse signal $\mathbf{\hat{S}}$ that is more focused on the MV area than for the RNMF based methods. The sparse signal of  NeuMF is also sparser on average than the one from the competing techniques, resulting in a higher MV segmentation performance on all echos with a satisfactory WD accuracy, and hence a higher MV segmentation performance on average.
\par
\begin{figure}[]
	\centering
	\begin{subfigure}{0.24\textwidth}
		\includegraphics[width=\textwidth]{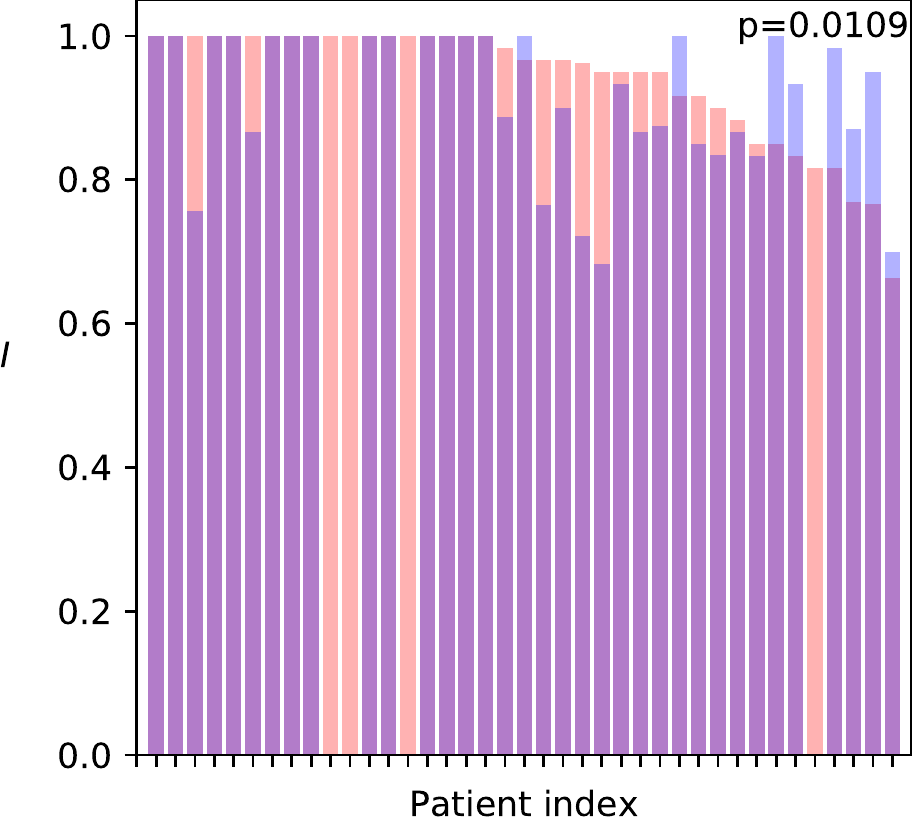}
	\end{subfigure}
	\begin{subfigure}{0.24\textwidth}
		\includegraphics[width=\textwidth]{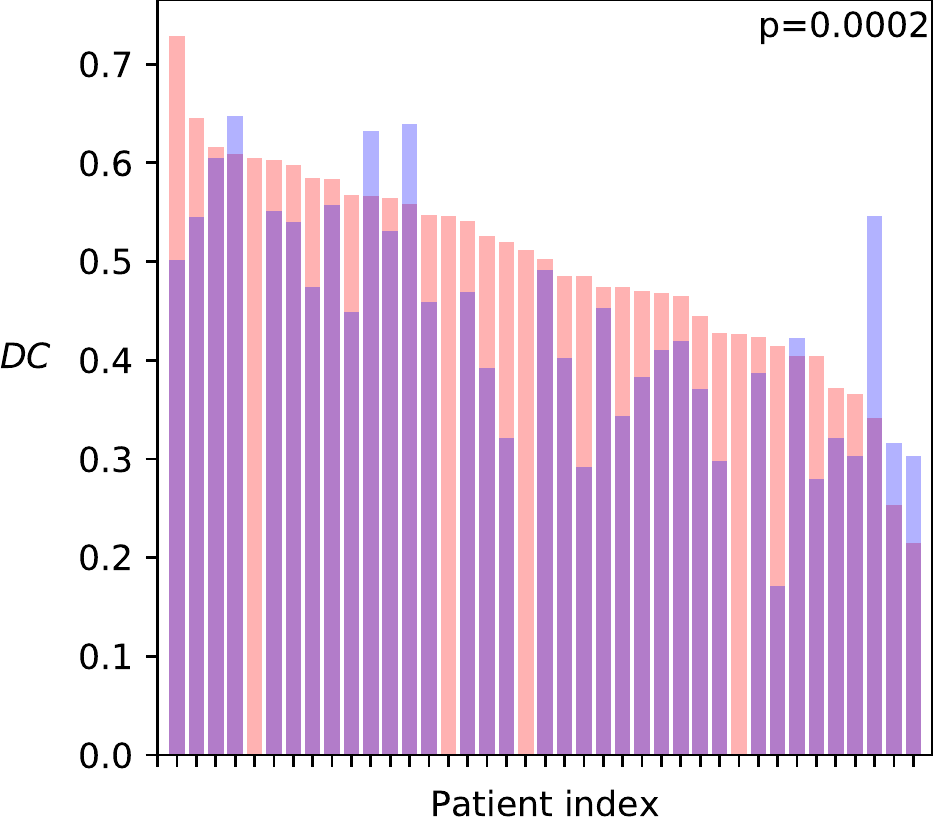}
	\end{subfigure}
	\begin{subfigure}{0.24\textwidth}
		\includegraphics[width=\textwidth]{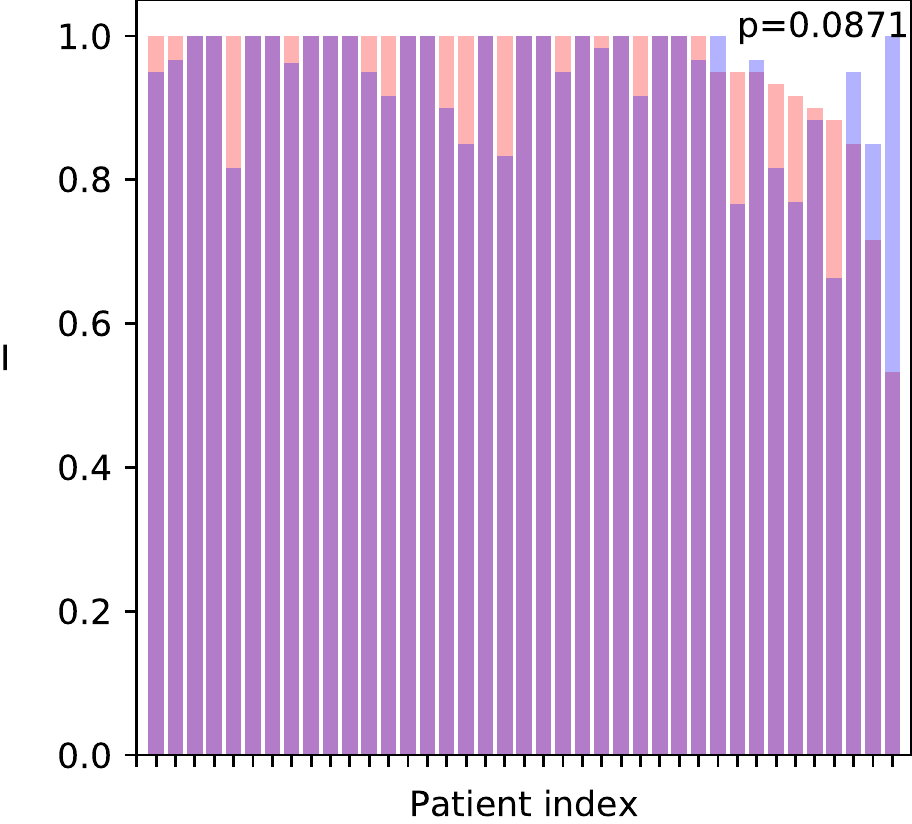}
	\end{subfigure}
	\begin{subfigure}{0.24\textwidth}
		\includegraphics[width=\textwidth]{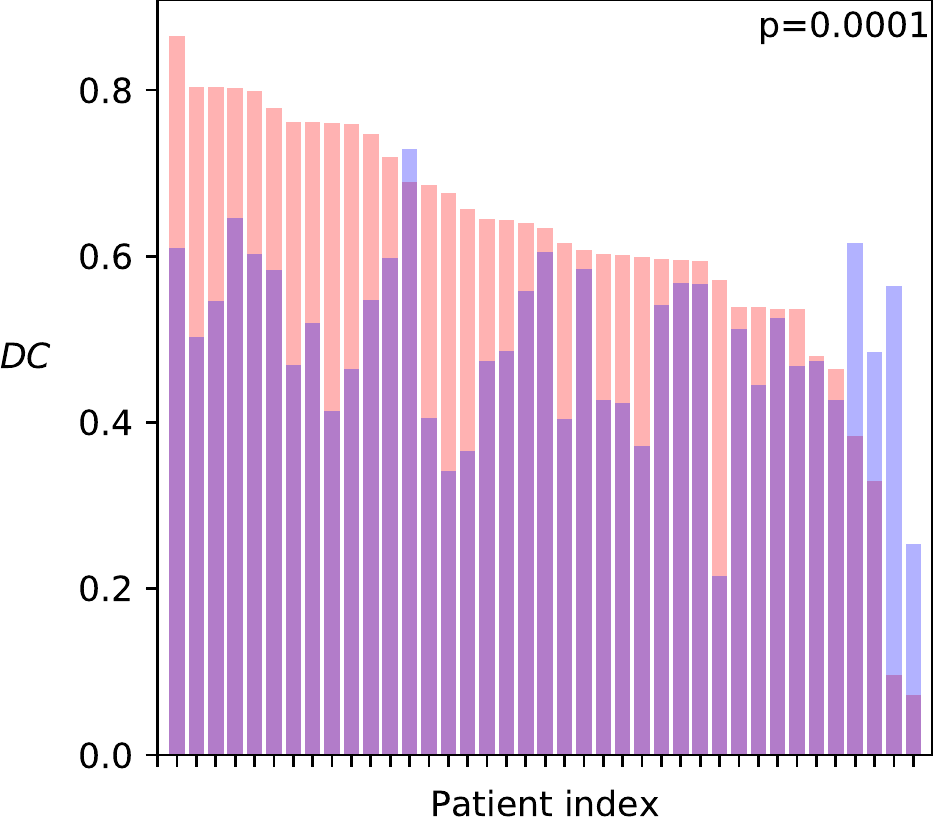}
	\end{subfigure}
	\caption{Performance on both the WD and MV segmentation tasks for the NN-MitralSeg, RNMF \cite{dukler2018automatic} and U-Net \cite{costa2019mitral} reported as in \Cref{fig:effect_time_masking}. The first two plots report the performance for NN-MitralSeg (red) and RNMF (blue) from left to right on WD and MV segmentation (scores sorted according to NN-MitralSeg). The last two columns report the performance for U-Net (red) and NN-MitralSeg (blue) from left to right on WD and MV segmentation (scores sorted according to U-Net). On the top right of each plot, the p-value for a one-sided t-test is reported.}
	\label{fig:nn_mitralseg_vs_literature}
\end{figure}

 In \Cref{tab:original_cohort_results}, we also observe that the supervised approach on average outperforms the NN-MitralSeg algorithm on IoU/Dice for the mitral valve segmentation with a high level of annotation. However, U-Net performance is highly dependent on the number of available training samples. With only two labels per echo, the supervised approach drops in segmentation quality below the NN-MitralSeg performance. Using only one labelled frame per echo impairs the predicted segmentation even in a stronger way. A detailed comparison of the NN-MitralSeg and the state-of-the-art of both unsupervised \cite{dukler2018automatic} and supervised \cite{ronneberger2015u,costa2019mitral} is provided in \Cref{fig:nn_mitralseg_vs_literature}.

\par
\begin{figure}[h]
\centering
\begin{subfigure}{0.5\columnwidth}
\includegraphics[width=0.95\textwidth]{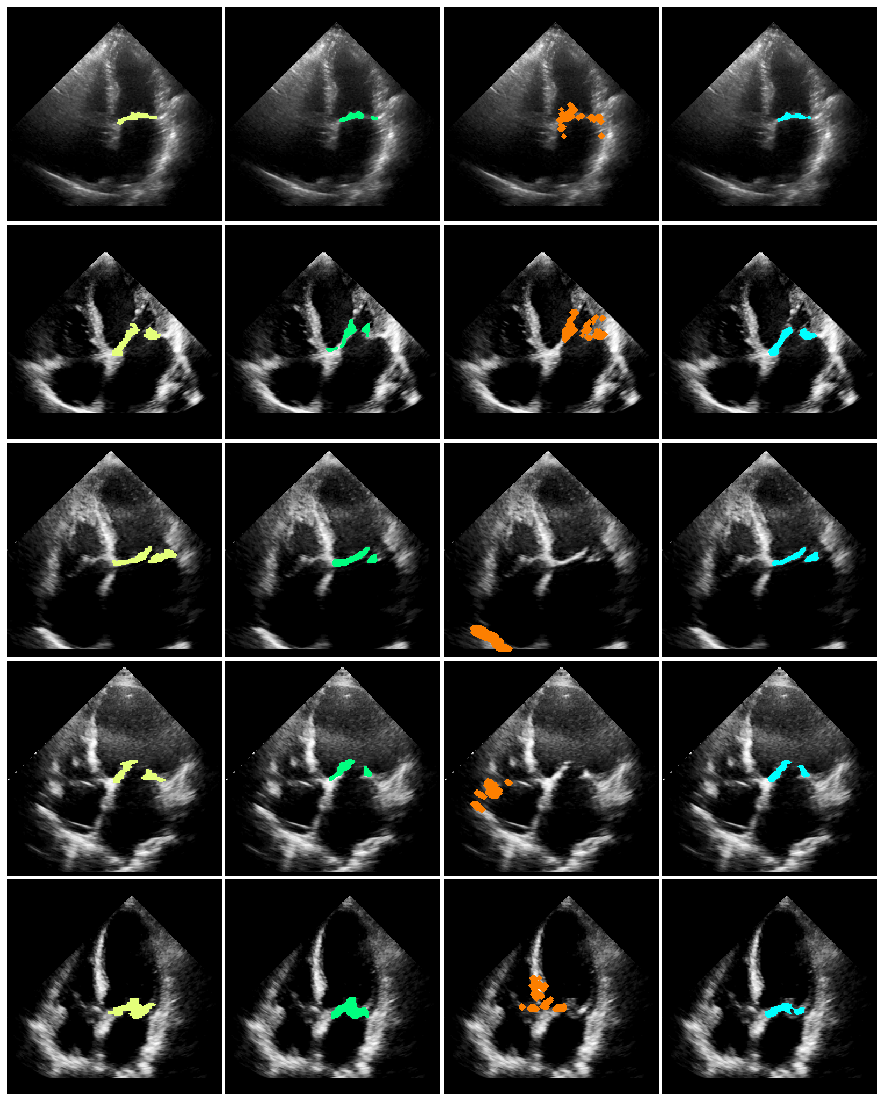}
\caption{}
\label{fig:top_5}
\end{subfigure}%
\begin{subfigure}{0.5\columnwidth}
\includegraphics[width=0.95\textwidth]{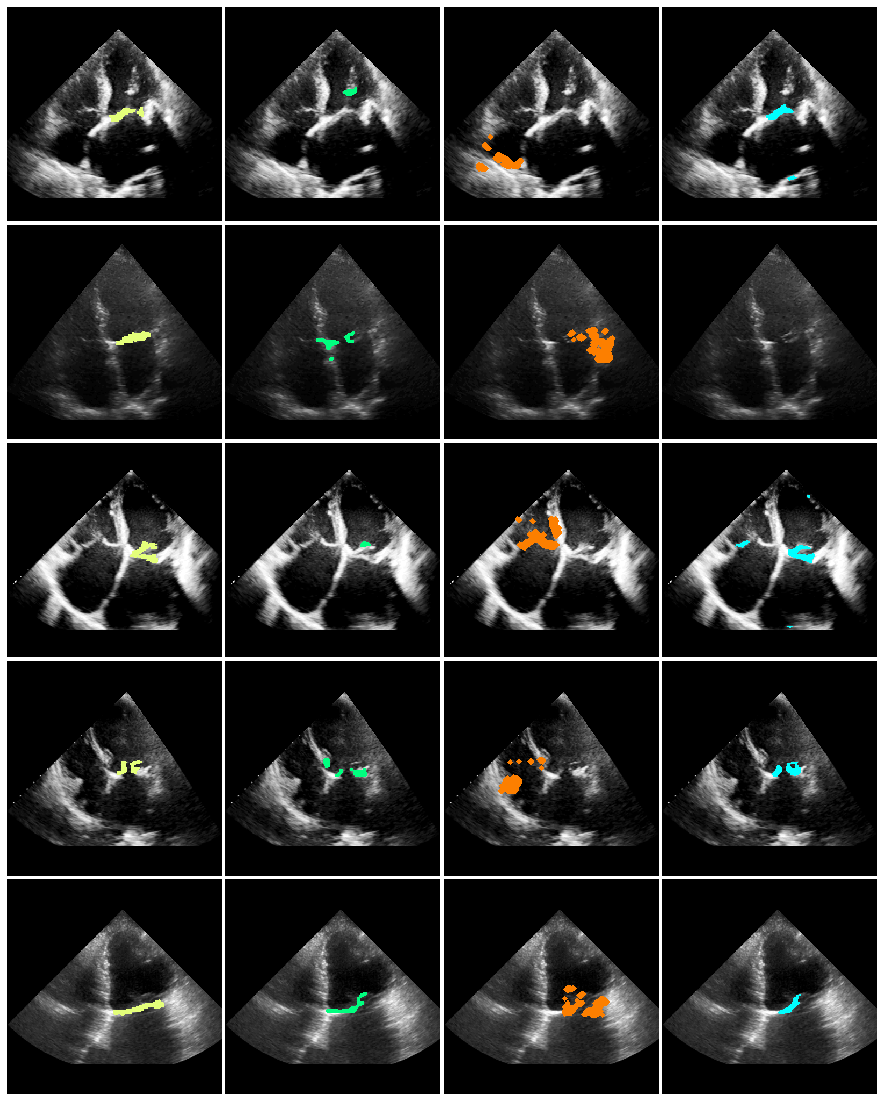}
\caption{}
\label{fig:bottom_5}
\end{subfigure}
\caption{The MV segmentation masks for the five echos with the (a) highest and (b) lowest Dice coefficients according to NN-MitralSeg. From left to right: ground truth (yellow), NN-MitralSeg (green) RNMF (red) and U-Net (blue). The WD algorithm for the RNMF approach is significantly less reliable than the one used in NN-MitralSeg. The U-Net performs better then the other method on average but fails on low-quality echos (second row, (b)) or if the tricuspid valve is clearly visible (third row (b)).}
\label{fig:seg_comparison_mask}
\end{figure}

\begin{figure}[h]
\centering
\includegraphics[width=\textwidth]{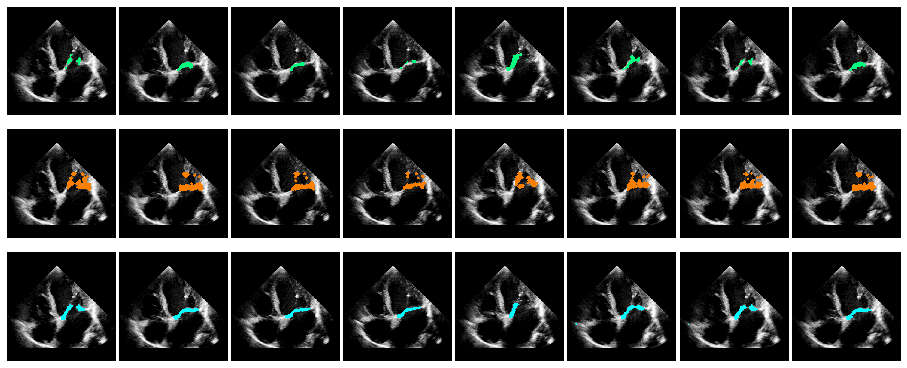}
\par \medskip
\includegraphics[width=\textwidth]{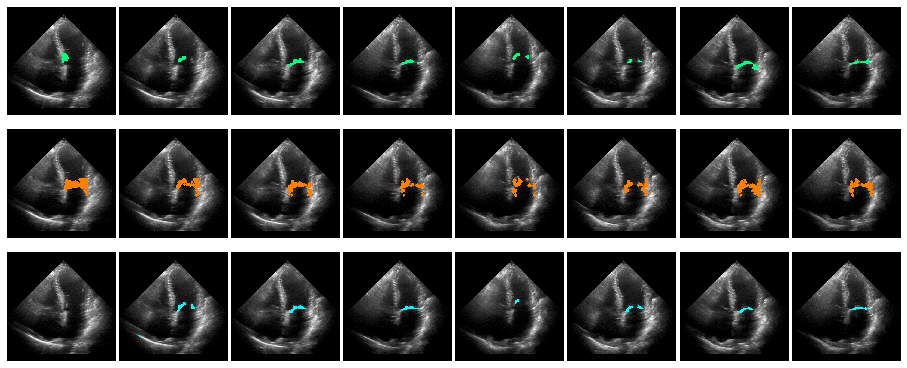}
\caption{The MV segmentation mask on two different echos (top three rows and bottom three rows) for eight frames separated by five time-steps each. NN-MitraSeg (green), R-NNMF (orange) and U-Net (blue).}
\label{fig:consecutive_frames}
\end{figure}

 A detailed comparison of the MV segmentation produced by NN-MitralSeg and the state-of-the-art RNMF \cite{dukler2018automatic} and U-Net \cite{ronneberger2015u} is documented in \Cref{fig:top_5,fig:bottom_5} and in \Cref{fig:consecutive_frames} where we show respectively the segmentation masks and the gold standard for the highest and lowest five scoring echos (according to our method), and the time consistency of the MV segmentation masks produced by the different methods.

\subsection{Post-processing}
\label{sec:post_processing}
\begin{figure}[h]
\centering
\includegraphics[width=0.8\textwidth]{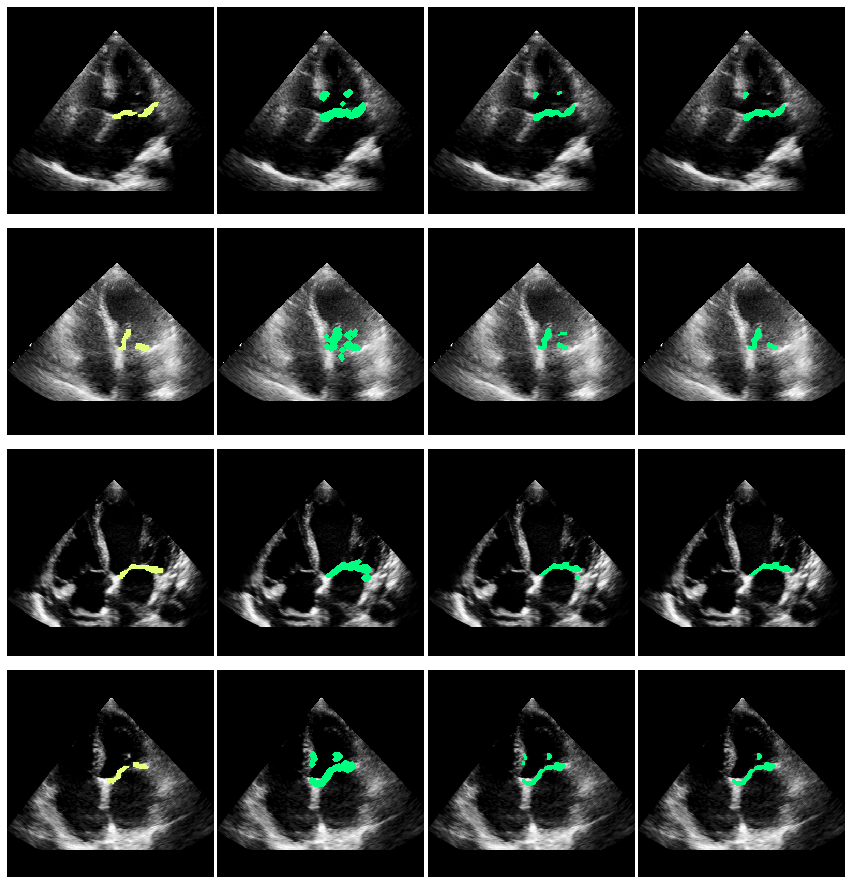}
\caption{Effect of the post-processing applied used in the NN-MitralSeg: The gold standard (yellow) and the segmentation (green) is depicted for four different echos. Columns from left to right report the valve segmentation mask respectively with no post-processing (second column), with the application of morphological operations (erosion and dilation, third column), and with the removal of the smallest connected components (calculated over the whole echo, last column).}
\label{fig:nnmf_post_processing}
\end{figure}

The effect of the post-processing steps used in the NN-MitralSeg algorithm is qualitatively depicted in \Cref{fig:nnmf_post_processing} and quantitatively reported in \Cref{tab:original_cohort_results}. Applying erosion and dilation confines the initial segmentation closer to the MV. Consequently, isolated parts of the segmentation masks are detected as the smallest connected components calculated over the complete echo and hence removed. Post-processing applied to the RNMF and NeuMF algorithms leads to better IoU and Dice scores. The IoU improves on average by 2.6\% for RNMF, by 2.1\% for NeuMF (RI) and by 3.8\% for NeuMF (MFI). The Dice coefficient increases on average by 1.7\% for RNMF, by 1.5\% for NeuMF (RI) and by 2.4\% for NeuMF (MFI).
The predicted segmentation masks by the U-Net are not improved by post-processing (see again \Cref{tab:original_cohort_results}, last three rows). The U-Net does not capture other fast-moving parts within the ROI, and segments only the parts belonging to the mitral valve. Applying morphological operations, i.e. erosion, reduces the size of the predicted area and hence decreases the segmentation performance.

\subsection{Computation issues and real-world deployment}
\begin{figure}[]
	\centering
	\begin{minipage}{0.98\textwidth}
	\begin{subfigure}[t]{0.36\textwidth}
		\centering
		\includegraphics[width=\textwidth]{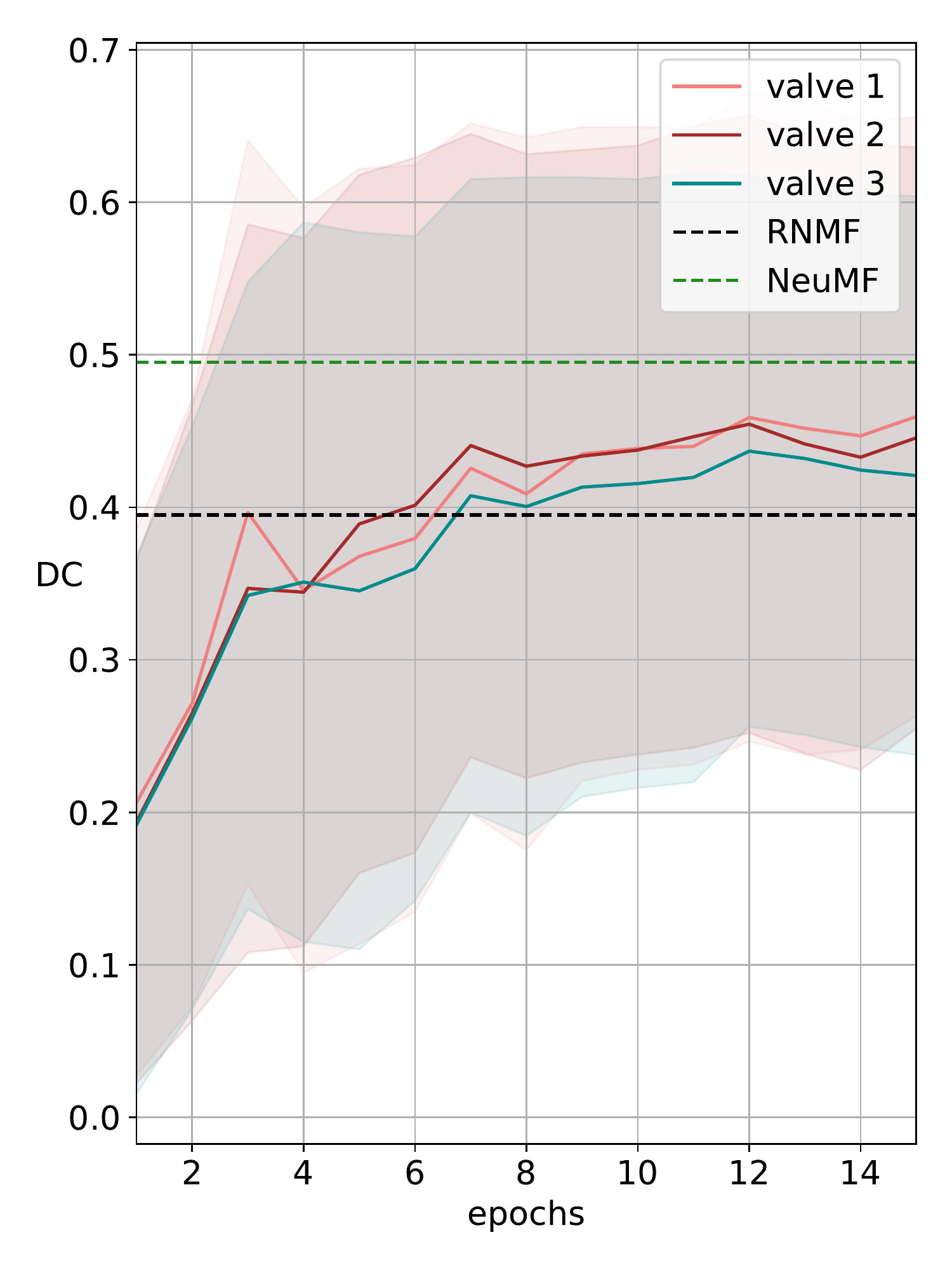}
		\caption{Average $DC$ computed on the three valves. The dashed lines are given for reference and depict the average $DC$ of the benchmarks RNMF and NN-MitralSeg trained on the full echos.}
		\label{fig:test_on_last_valve_scores}
	\end{subfigure}
	\quad
	\begin{subfigure}[t]{0.60\textwidth}
		\centering
		\includegraphics[width=\textwidth]{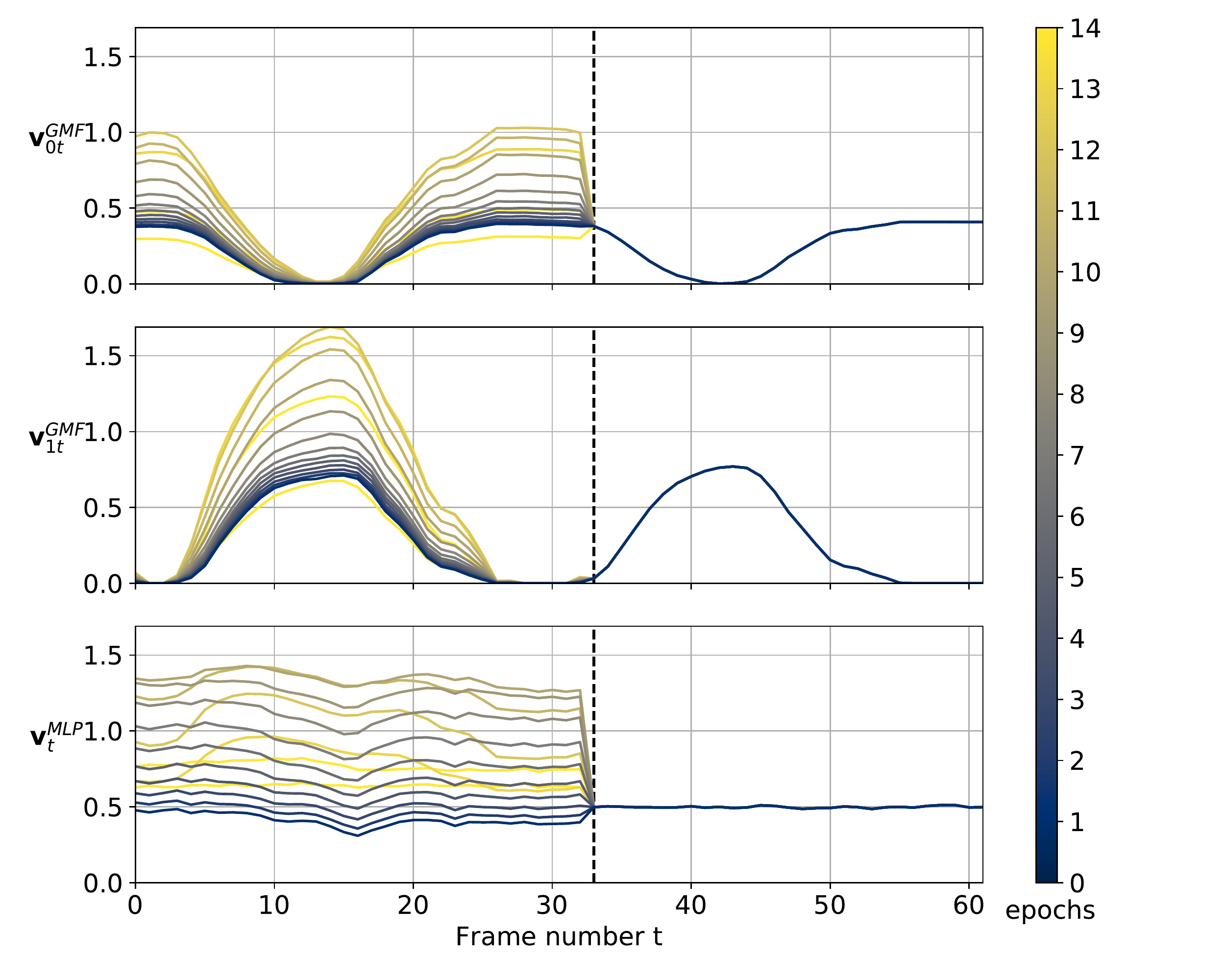}
		\caption{Temporal embedding vectors $\mathbf{v}^{GMF}$ and $\mathbf{v}^{MLP}$ during training on a single echo. The training is performed only on the first section of the echo, depicted by the dashed vertical line. The trained model is then deployed in the second part of the video with no further training.}
		\label{fig:test_on_last_valve_time_series}
	\end{subfigure}
	\end{minipage}
	\caption{Experiment of the deployment of the pretrained NN-MitralSeg method on new echos. Full details in the main text.}
	\label{fig:test_on_last_valve}
\end{figure}
The method so far described, together with all the other unsupervised methods based on low dimensional factorization method and subsequent outlier detection, requires the training of the full model (neural network and spatial-temporal embeddings) on every new echocardiography video available. In our experiments, this task required an average of 13 $\pm$ 8 min per video on a single GeForce GTX TITAN X GPU. Despite this computational burden being limited, it can result in a barrier in the clinical practice, especially in time-critical scenario like in intensive care units or operative procedure. For this reason, in the following, we design an experiment that shows the viability of the method in these circumstances. The echos are split into two consecutive parts such that the two parts contain respectively two and one labelled MVs. The NN-MitralSeg model is then trained only on the first part of the echo (with two labelled MVs) and tested on the second part. This setting simulates the scenario in which an echo has been processed by the NN-MitralSeg method, and subsequentially the method is deployed on a new echo of the same patient, hence the neural network weights and all the spatial embeddings can be redeployed with no further training. In \Cref{fig:test_on_last_valve_time_series}, we can observe the qualitative behaviour of the time-variant embedding during training, with the embedding vector partitioned into a trained and a frozen section. The performance of the method is reported in \Cref{fig:test_on_last_valve_scores}, with the individual dice scores for the three valves, averaged on the whole dataset. We can observe that despite the model being trained only on one part of the echo, the segmentation performance on the last valve increases during training, hence the model can generalize well to the section of echos that has not been trained on. We can further observe that despite the model performing on average worse on the last valve than on the other two, it can outperform the RNMF method (dashed black line) trained on the full echos. The drop in the dice coefficient from the fully trained model (green dashed line) is due to the decreased amount of trained data and amounts in $10\%$ decrease on average. 
\section{Conclusion and future work}
\label{sec:conclusion}
We proposed NN-MitralSeg, a fully automated and unsupervised mitral valve segmentation algorithm based on non-linear matrix factorization using neural network collaborative filtering. An echocardiography video is decomposed into a low dimensional signal that captures the linear and non-linear myocardial wall motion, and a high dimensional sparse signal that accounts for the echocardiography noise and mitral valve movement. The mitral valve is then segmented from the sparse signal using thresholding, diffusion algorithms and morphological operations. This method outperforms the state-of-the-art fully automated unsupervised algorithm in a data-set of 39 videos with patients suffering various mitral valve dysfunctions and in a independent public data-set, in both the task of positioning the rectangular region of interest and in the accuracy of the dense mitral valve mask. NN-MitralSeg also compares favourably with the state-of-the-art supervised method, (i) outperforming it on a low level of annotation and (ii) being more robust to low-quality echo at every level on annotation considered. These performance benefits demonstrate that prior assumptions of the structure of echocardiography videos are of great importance in the tasks considered. Furthermore, while NN-MitralSeg could easily be extended to 2CH or 3CH echo views, the supervised approach would most likely fail if not retrained again with new labels. 
\par Despite the modest size of the dataset, the reported experiments contain echo videos with a larger variability than a healthy control dataset of the same size, thereby documenting the robustness of the method. Possible future developments include the use of both sparse ground truth segmentation masks and dense (inaccurate) annotation generated by unsupervised algorithms (like NN-MitralSeg) to train deep networks for segmentation in a weakly-supervised learning scenario \cite{zhou2017brief}. To ensure relevance in clinical practice, the segmentation algorithms should also be extended to efficient online processing. 
Enforcing a factorization-like structure in the embedding space of the U-Net architecture can be expected to retain the best of both approaches. This design choice would provide practitioners with segmentation algorithms that could be deployed in real-time echocardiography during mitral valve intraoperative procedures.
\section*{Acknowledgments}
Joachim M. Buhmann has been supported by PHRT (ETH) for the SWISSHEART Failure Network.

\newpage
\bibliographystyle{unsrt}
\bibliography{bibliography}
\newpage

\begin{appendices}
\crefalias{section}{appsec}
\section{Results EchoNet-Dynamic dataset}
\label{app:echonet}
We evaluated our segmentation algorithm on the publicly available echocardiographic video dataset (EchoNet-Dynamic Dataset \cite{ouyangechonet}). 46 videos were selected and for each video three frames were labelled by medical experts. The resolution of the videos is only 112x112 pixels. Despite the low resolution, the NN-MitralSeg manages to achieve good results (see \Cref{tab:echonet_results}). However, due to the low resolution of the videos, using optical flow (OF) for window detection impairs the performance. In \Cref{tab:echonet_results}, we compare our algorithm against the previous state of the art unsupervised segmentation algorithm based on robust non-negative matrix factorization (RNMF) \cite{dukler2018automatic} and against the supervised U-Net \cite{ronneberger2015u,costa2019mitral}.
\begin{table}[h]
\caption{Summary of the results for both window detection and mitral valve segmentation methods evaluated on the EchoNet-Dynamic dataset. For the unsupervised method RNMF and NeuMF, the post-process refinement introduced in \Cref{sec:post_processing} is used. Full details in the caption of \Cref{tab:original_cohort_results}.}
\centering
\small
\setlength\tabcolsep{4pt}
\begin{tabular}{cc c | c c c c | c c c c}
\toprule
\textbf{Method}& \bfseries WD & \bfseries TM & $\bm{I_{65}}$ & $\bm{I_{85}}$ & $\bm{IoU}$ & $\bm{I}$ & $\bm{IoU}$ & $\bm{DC}$\\
\midrule
\multirow{1}{*}{\textbf{RNMF}} & TO & \crossed & 36  & 30  &  0.367 & 0.790  & 0.197 & 0.314\\
\multirow{4}{*}{\shortstack{\textbf{NeuMF} \\ \textbf{MFI}}} & TO & \crossed & 41  & 40  & 0.422  & 0.887  &  0.321 & 0.461 \\
& TO & \checkmark & \bfseries 45 & \bfseries 42 & \bfseries 0.451 & \bfseries 0.957 & \bfseries 0.375 & \bfseries 0.531\\
& OF & \crossed & 23 & 12 & 0.229 & 0.517 & 0.159 & 0.239\\ 
& OF & \checkmark & 29 & 16 & 0.298 & 0.668 &  0.202 &  0.302\\
\midrule
& & $\bm{l}$ & & & & & & \\
\midrule
\multirow{1}{*}{\textbf{U-Net}}  &  & 3 & \multirow{1}{*}{46} & \multirow{1}{*}{41} & \multirow{1}{*}{ 0.447} & \multirow{1}{*}{ 0.946} & 0.515 & 0.660 \\

\bottomrule
\end{tabular}
\label{tab:echonet_results}
\end{table}

\section{Additional visualizations}
\label{app:embeddings}
\begin{figure}[h]
\centering
\begin{subfigure}{0.8\textwidth}
\includegraphics[width=\textwidth]{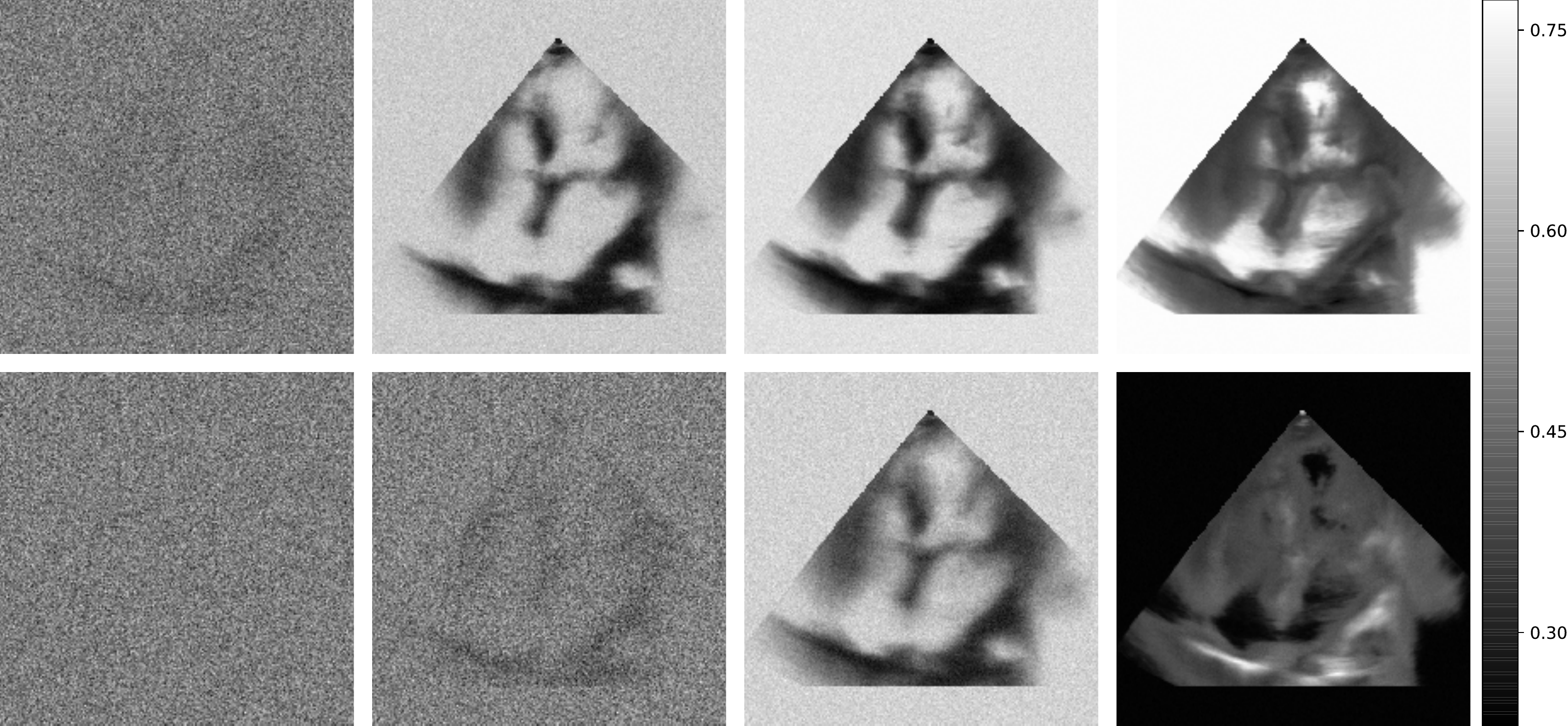}
\caption{Embedding vectors $\mathbf{u}^{MLP}_n$.}
\label{fig:mlp_u}
\end{subfigure}
\begin{subfigure}{0.8\textwidth}
\includegraphics[width=\textwidth]{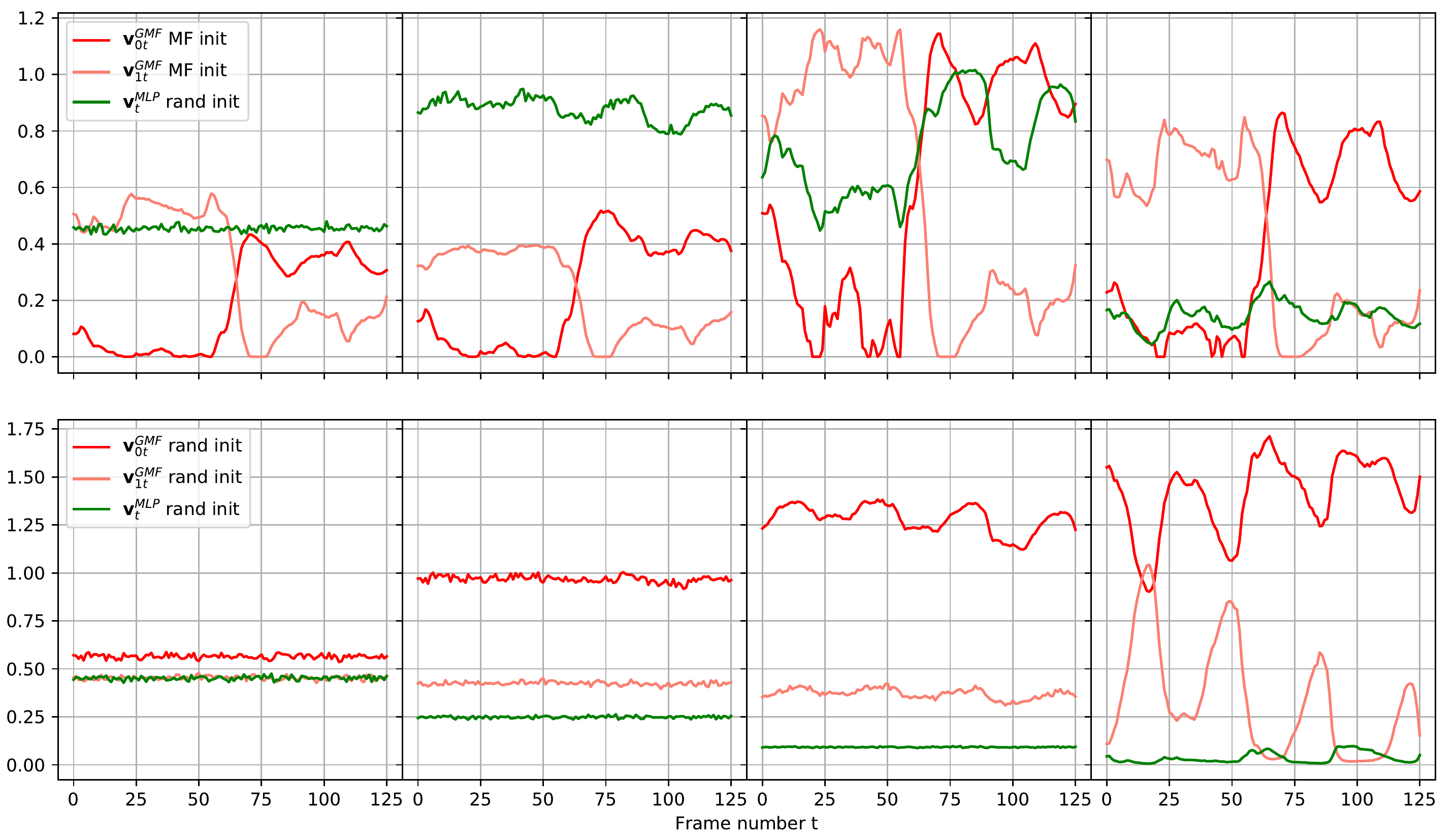}
\caption{Emdedding vectors $\mathbf{v}^{GMF}_n$ and $\mathbf{v}^{MLP}_n$.}
\label{fig:gmf_v}
\end{subfigure}
\caption{Emdedding vectors during training, for both MFI (top) and RI (bottom) at the four step indicated by the dashed vertical lines in \Cref{fig:training_curves}. In (a) the images are built as in \Cref{fig:gmfu_mfi}. For both cases the embedding dimensions are $K=2$ and $K'=1$.}
\end{figure}

In \Cref{fig:mlp_u}, we report the embedding vector $\mathbf{u}^{MLP}_n$ for the two different NeuMF initialization (RI and MFI) on one echo in four sequential stages of training. We can observe that the model embeds a frame that resembles an average frame, being more blurred and less defined on the myocardium walls, and hence accounting for a bias term of the GMF column of the model. We can also notice that the MFI is faster in building a spatial structure in the embedding vector.

\begin{figure}
\centering
\includegraphics[width=0.35\textwidth]{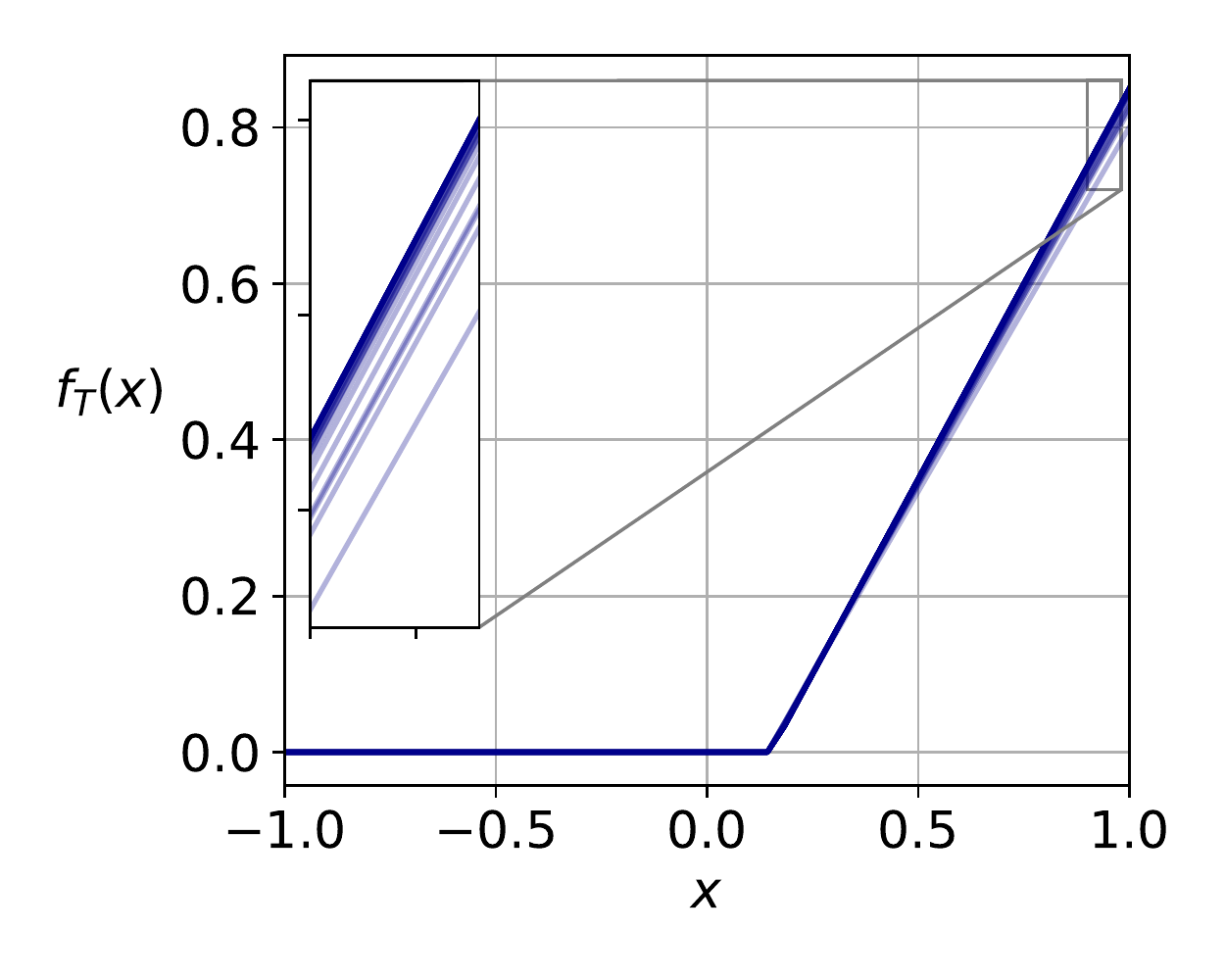}
\caption{Threshold network values in the interval $\{-1,1\}$ and the end of the training for all the echos considered. }
\label{fig:thresholds}
\end{figure}

In \Cref{fig:gmf_v}, we can observe the time-variant embedding vectors $\mathbf{v}^{GMF}_t$ and $\mathbf{v}^{MLP}_t$. Note that the MLP embedding vector $\mathbf{v}^{MLP}_t$ is randomly initialized in both cases, and it gains a complex time-structure in the MFI while being close to zero and with little time structure in the RI.
In \Cref{fig:thresholds} we also report the thresholding function values learned by the network over all echos considered. The thresholding function changes only sightly between different echos and behaves indeed as a threshold operator.

\section{Model specifications for NeuMF}
\label{app:net_specs}
Across both datasets (MitraSwiss registry and EchoNet) and all echos, the hyperparameters of the model are kept fixed as follow. The Adam optimizer \cite{kingma2014adam} is used with a learning rate of $0.005$ and a batch size of $10,000$. The dimensions of the latent features are kept constant across all echos at $K=2$, and $K'=1$. The sparsity coefficient and regularization parameter are also kept constant across all videos at $\lambda=0.3$ and $\beta=0.1$.
The window size used for the window detection procedure is fixed to $60\times80$. The standard deviation of the Gaussian to smooth the temporal derivatives in the optical flow algorithm is set to 3.5.
Both networks (low dimensional and threshold) consists of three fully connected (FC) layers with 10 units each. The latent features and the first two FC layers have ReLU  and the last FC layer has sigmoid activation to give $\hat{X}_{n,t}\in [0,1]$. The sofplus activation function is used to obtain non-negative embedding vectors. 
Xavier initialization is performed with gain one, while all the Gaussian initializations are performed on the embedding vectors with average 0.5 and standard deviation 0.01. The number of training epochs is held constant among all echos and is fixed to 15. 

\section{U-Net implementation}
\label{app:unet_specs}
The network is fully convolutional and consists in total of 18 convolutional layers with a kernel size of 3x3 and ReLU activation function, and a 1x1 convolutional layer with a sigmoid activation function for the output layer. The encoder consists of four max pooling operations (2x2) which are applied after every second convolutional layer. In the decoder part, up-convolutions (transposed convolutions) with a kernel size of 2x2 and strides of (2, 2) are used. For the mitral valve segmentation and the ROI selection, the same U-Net architecture described above is used. The Adam optimizer \cite{kingma2014adam} is used with a learning rate of $0.001$, $\beta_1=0.9$ and $\beta_2=0.999$. The binary cross-entropy loss\footnote{Applying weighted binary cross-entropy to account for the highly imbalanced class distribution, i.e. valve vs. background, did not lead to better results.} is used for training. A batch-size of 16 is used for both the mitral valve segmentation and the ROI selection. The models are trained with early stopping with the validation loss evaluated on 10\% of the training. The models' performance are evaluated with 13-fold cross-validation. The splitting is performed over videos (36 videos for training and 3 videos for testing) for every cross-validation fold, and the average of the testing performance is reported in both \Cref{tab:original_cohort_results} and \Cref{tab:echonet_results}. For the MV segmentation either one, two or three labelled frames per video are used, while for the ROI prediction one label for each frame (one ground-truth ROI per video) is available.
The final mitral segmentation is achieved by thresholding the output of the networks at $0.5$. The final ROI selection is determined by averaging the ROI predictions of individual frames of one echo, calculating the centre point of the averaged prediction and then taking a fixed window size of $80 \times 60$ pixel around this centre point. This is done to ensure a fair comparison with the NN-MitralSeg window detection algorithm.
\section{Segmentation with the active contour method}
\label{app:active_contour}
\begin{figure}[h]
	\centering
	\begin{subfigure}{0.80\textwidth}
		\includegraphics[width=1\linewidth]{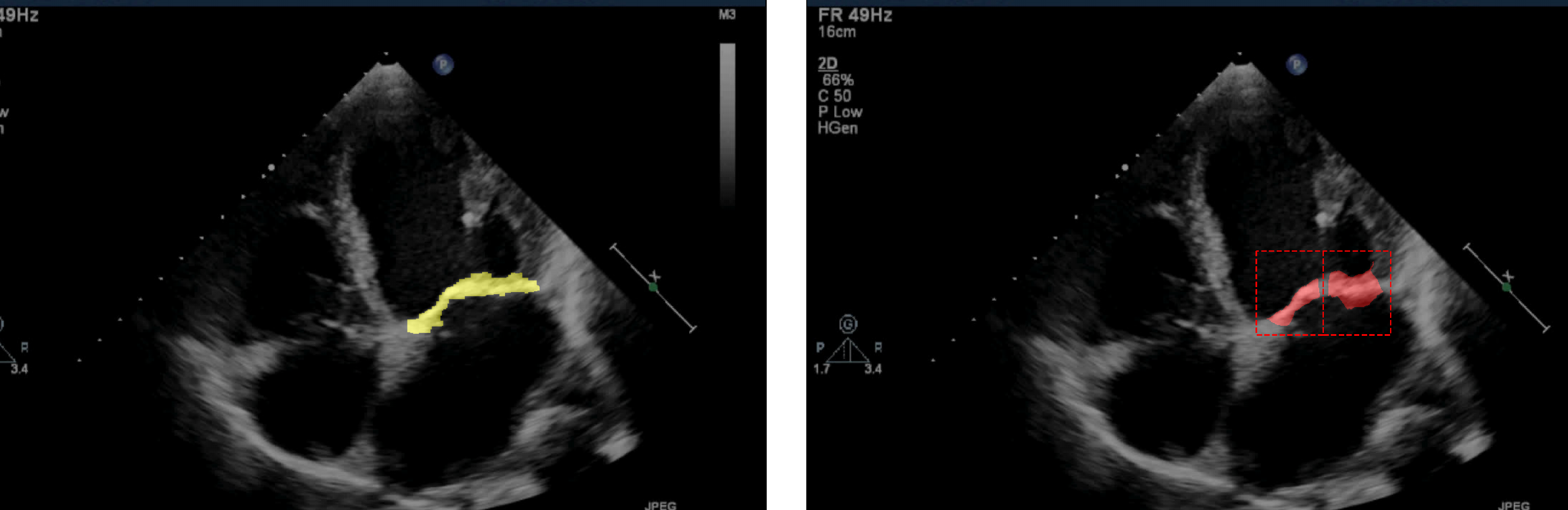}
	\end{subfigure}\\
	\vspace{0.2cm}
	\begin{subfigure}{0.80\textwidth}
		\includegraphics[width=1\linewidth]{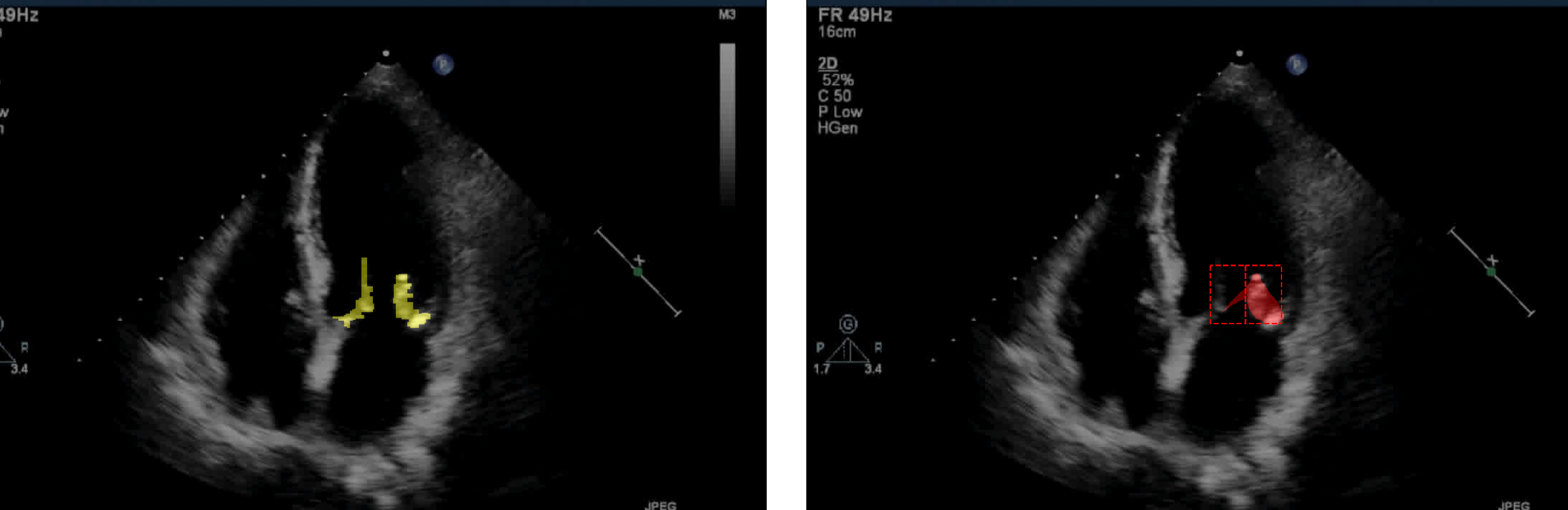}
	\end{subfigure}
	\caption{Top row: Example of the active contour method success in delineating the mitral-valve. Bottom row: Example where the active contour method fails to delineate one of the leaflets due to a bright spot in the other (right) leaflet (yellow: ground-truth, red: AC prediction).}
	\label{fig:ac_good_bad}
\end{figure}

For the implementation of the active contour method (AC) \cite{kass1988snakes} we use the scikit-image library for image processing \cite{van2014scikit}. The AC method uses a spline to define the contour of the MV to be segmented and minimizes an energy functional that is in part defined by the image (so that the spline follows contours) and part by the spline’s shape, i.e. length and smoothness (so that the splines cannot be arbitrary rugged). Active contour methods require good manual initialization to achieve acceptable performance. In case the bounding box used for the initialization is too large the method fails to segment the mitral valve accurately. The main reason is that the myocardium close to the MV provides sharper intensity gradients then the MV itself. In \Cref{fig:ac_good_bad} we provide two selected examples of the predicted segmentation of the AC against the ground truth label in case of success (top row) and failure (bottom row) of the method. 

\begin{figure}[H]
	\centering
	\begin{subfigure}{\textwidth}
		\centering
		\includegraphics[width=0.48\textwidth,height=1.55cm]{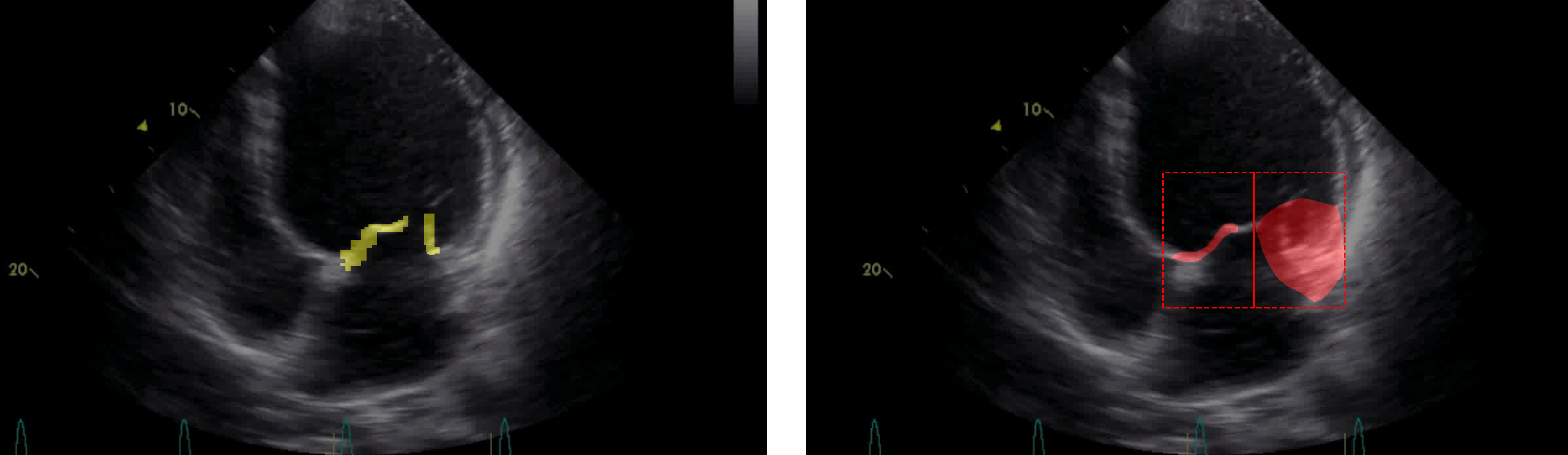}
		\hfill
		\includegraphics[width=0.48\textwidth,height=1.55cm]{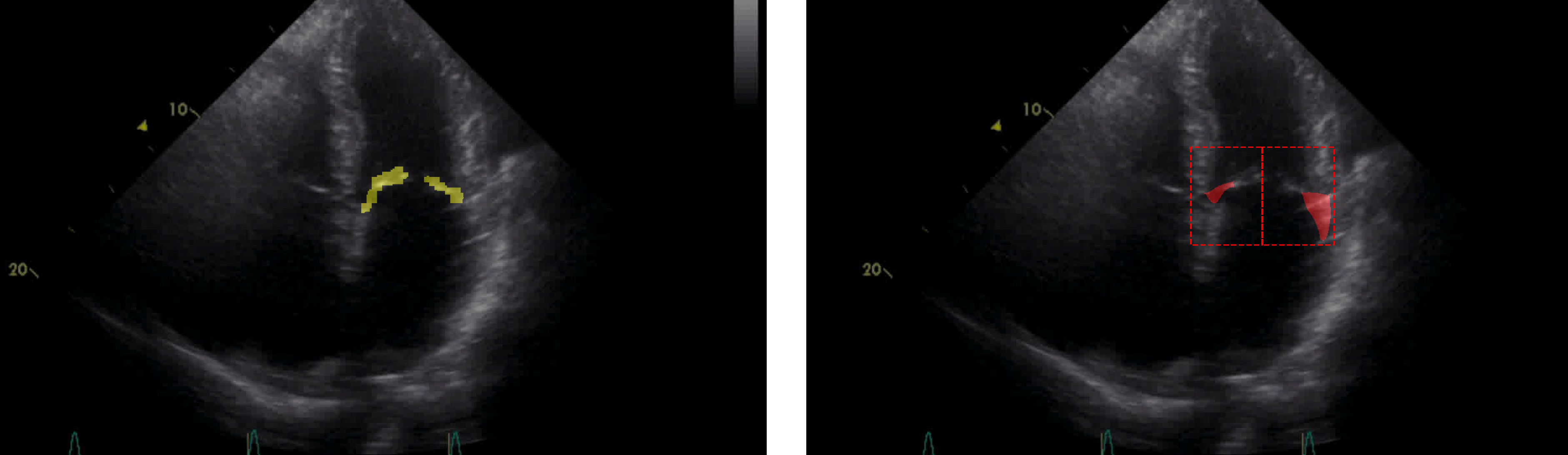} \\\vspace{0.1cm}
		\includegraphics[width=0.48\textwidth,height=1.55cm]{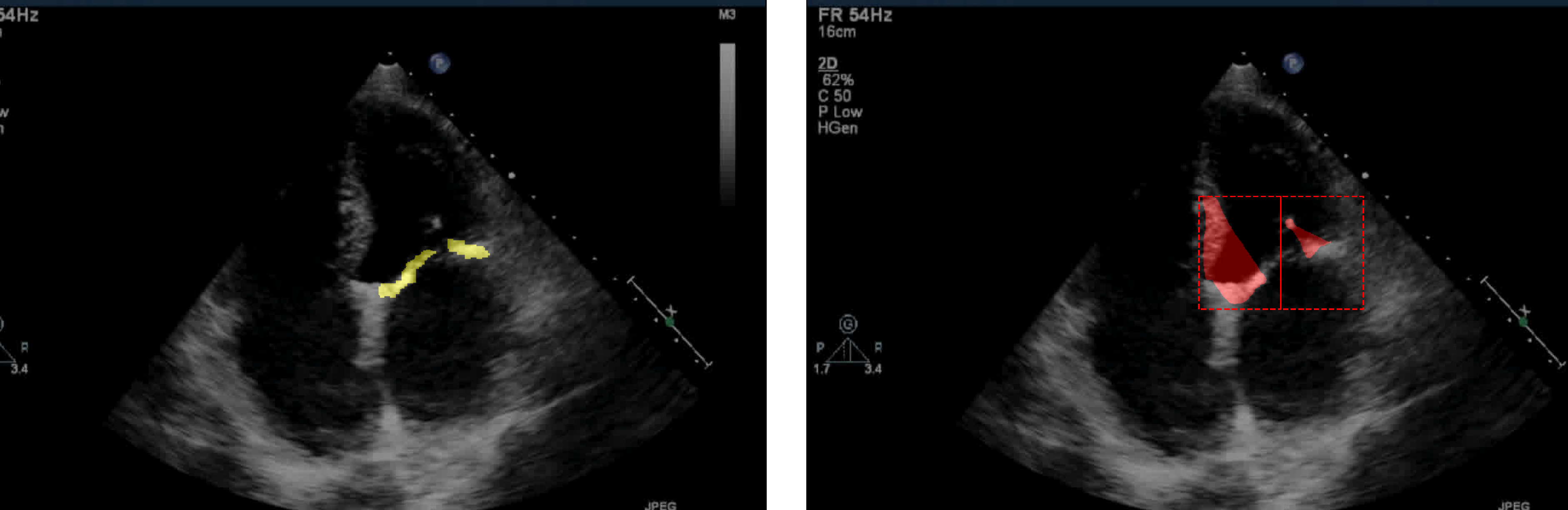}
		\hfill
		\includegraphics[width=0.48\textwidth,height=1.55cm]{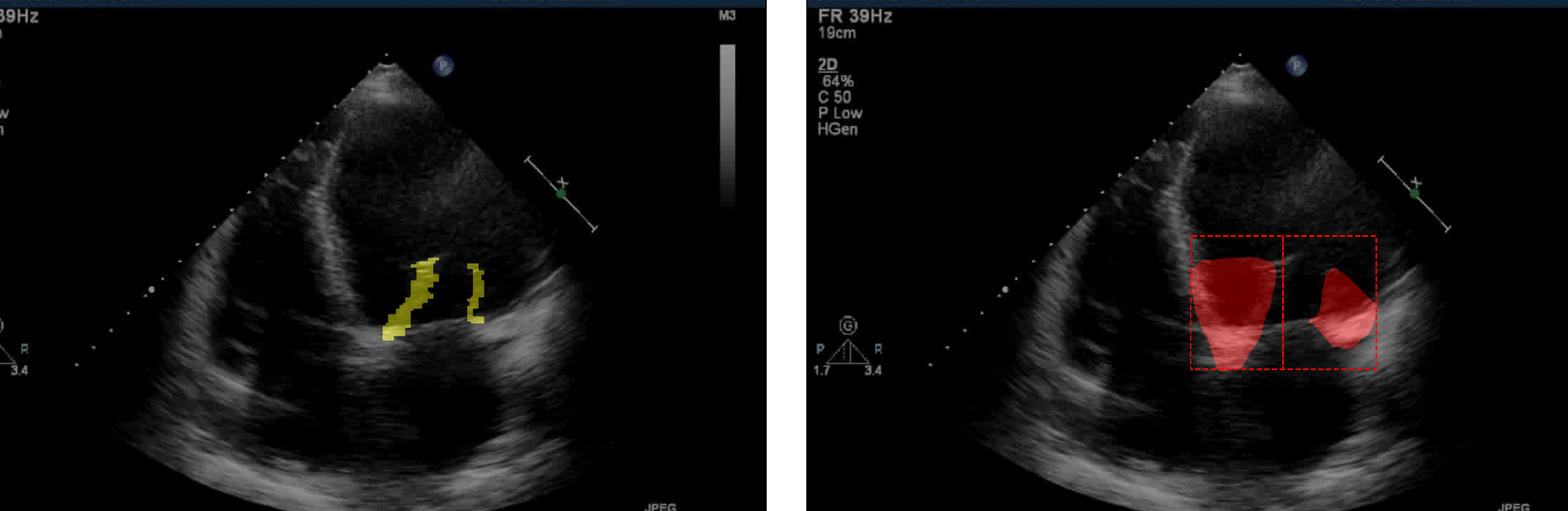} \\\vspace{0.1cm}
		\includegraphics[width=0.48\textwidth,height=1.55cm]{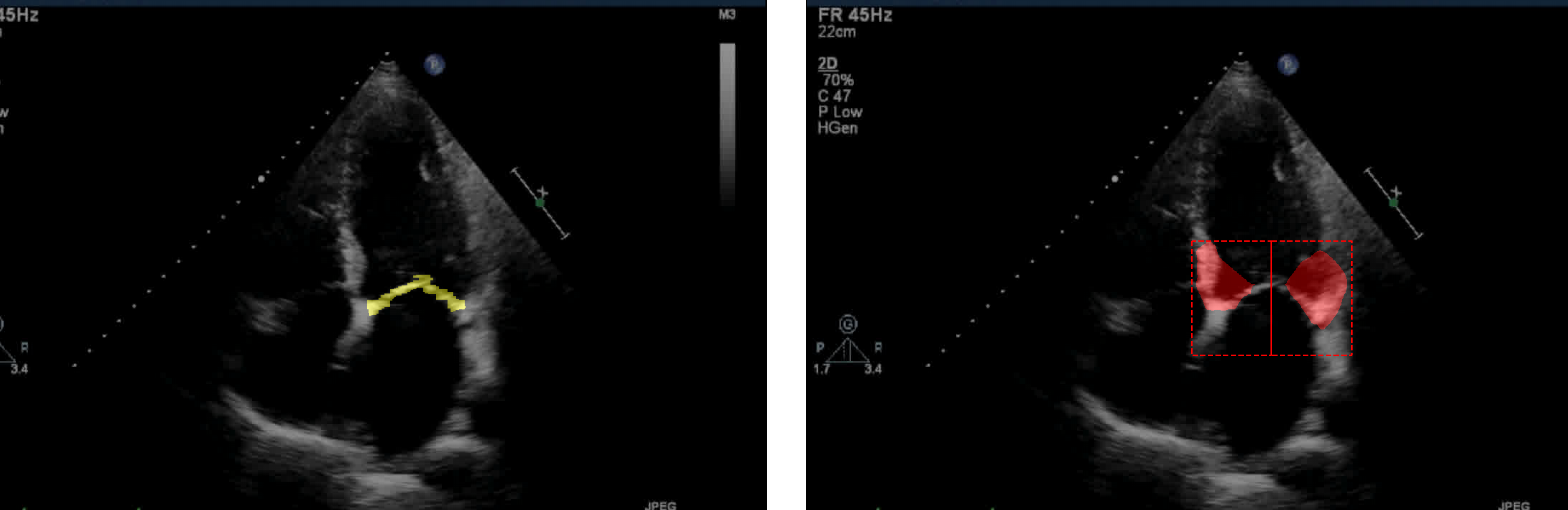}
		\hfill
		\includegraphics[width=0.48\textwidth,height=1.55cm]{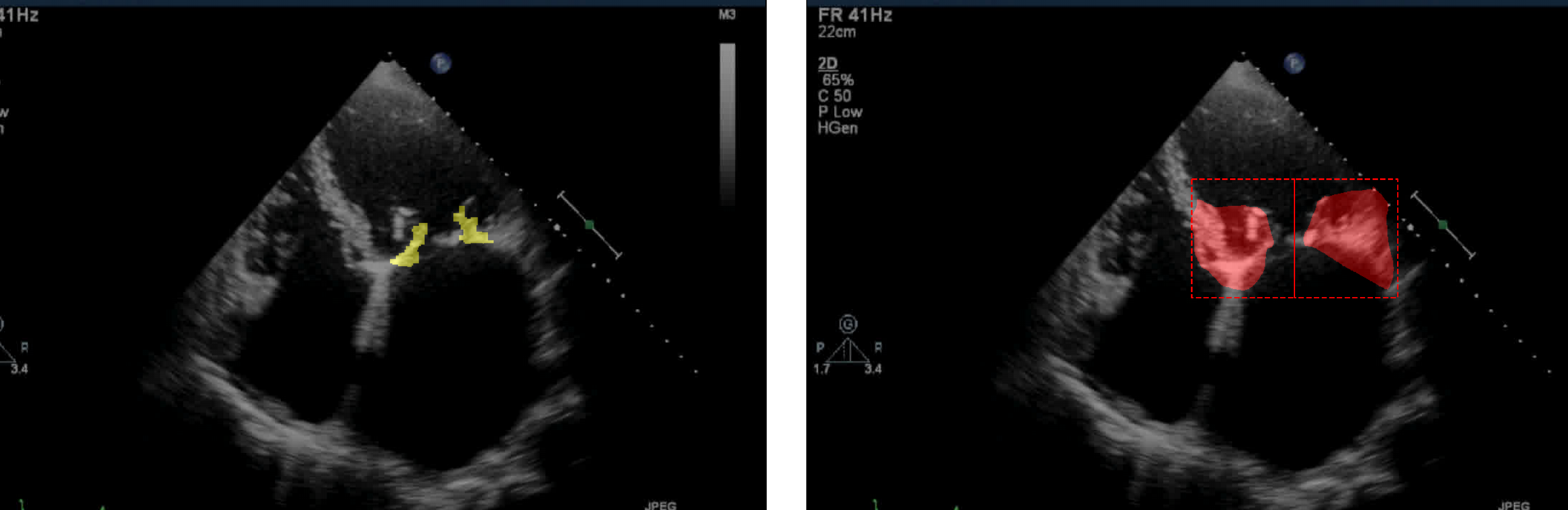} \\\vspace{0.1cm}
		\includegraphics[width=0.48\textwidth,height=1.55cm]{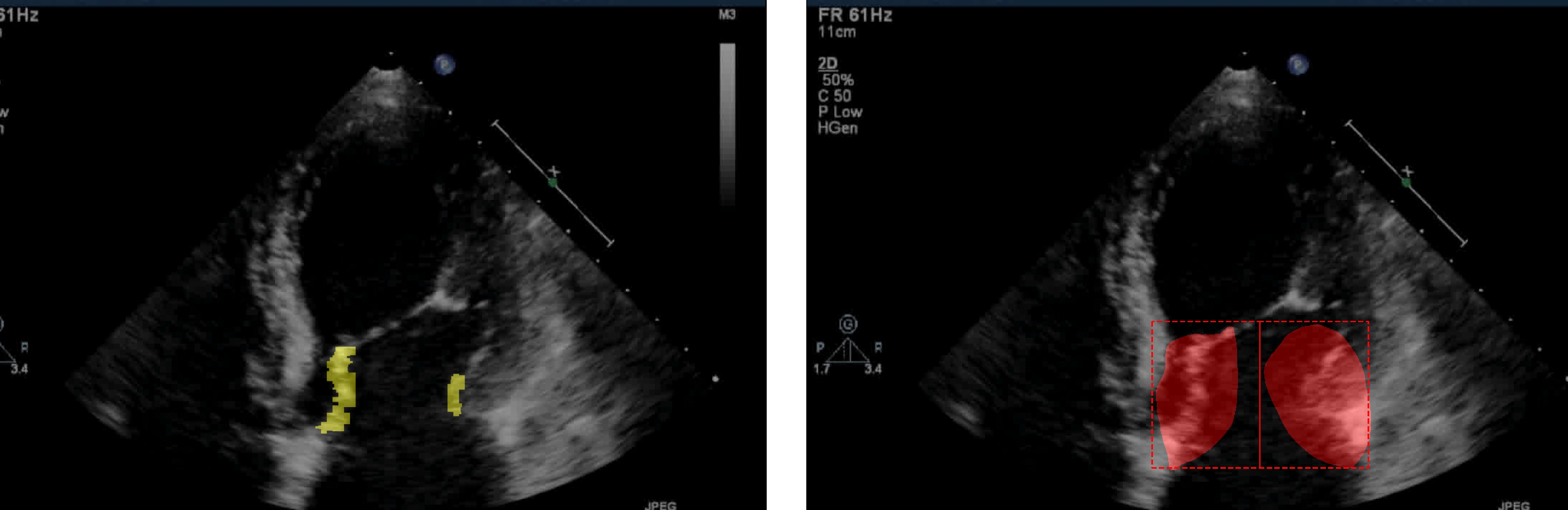}
		\hfill
		\includegraphics[width=0.48\textwidth,height=1.55cm]{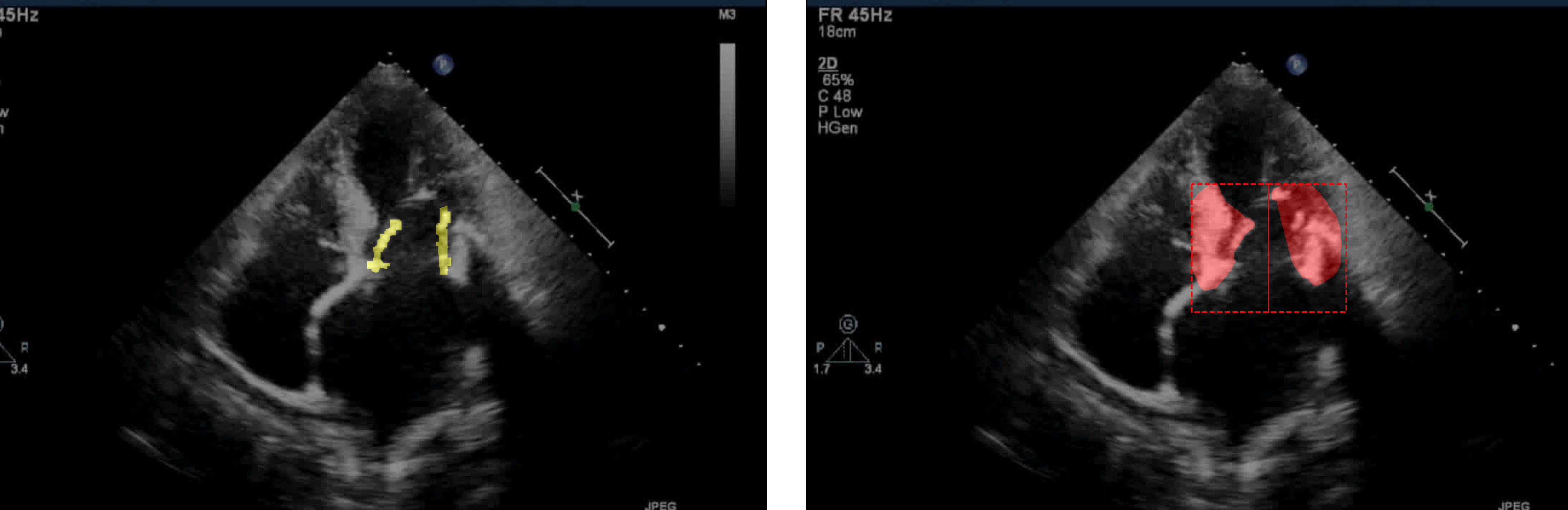} \\\vspace{0.1cm}
		\includegraphics[width=0.48\textwidth,height=1.55cm]{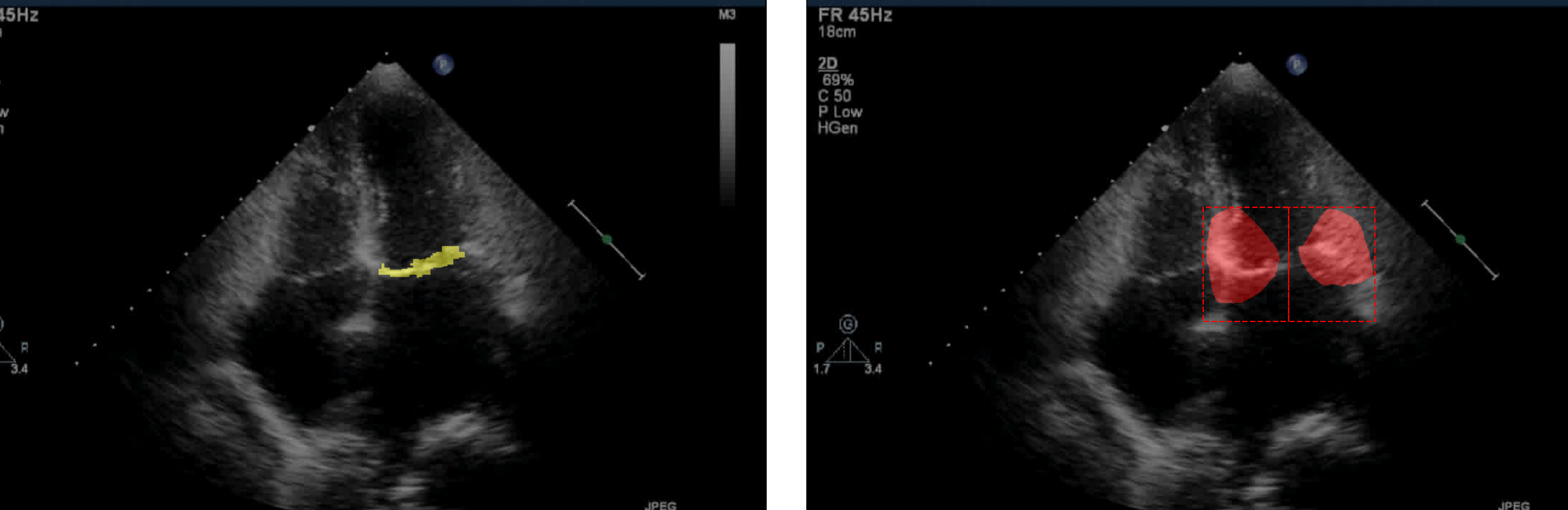}
		\hfill
		\includegraphics[width=0.48\textwidth,height=1.55cm]{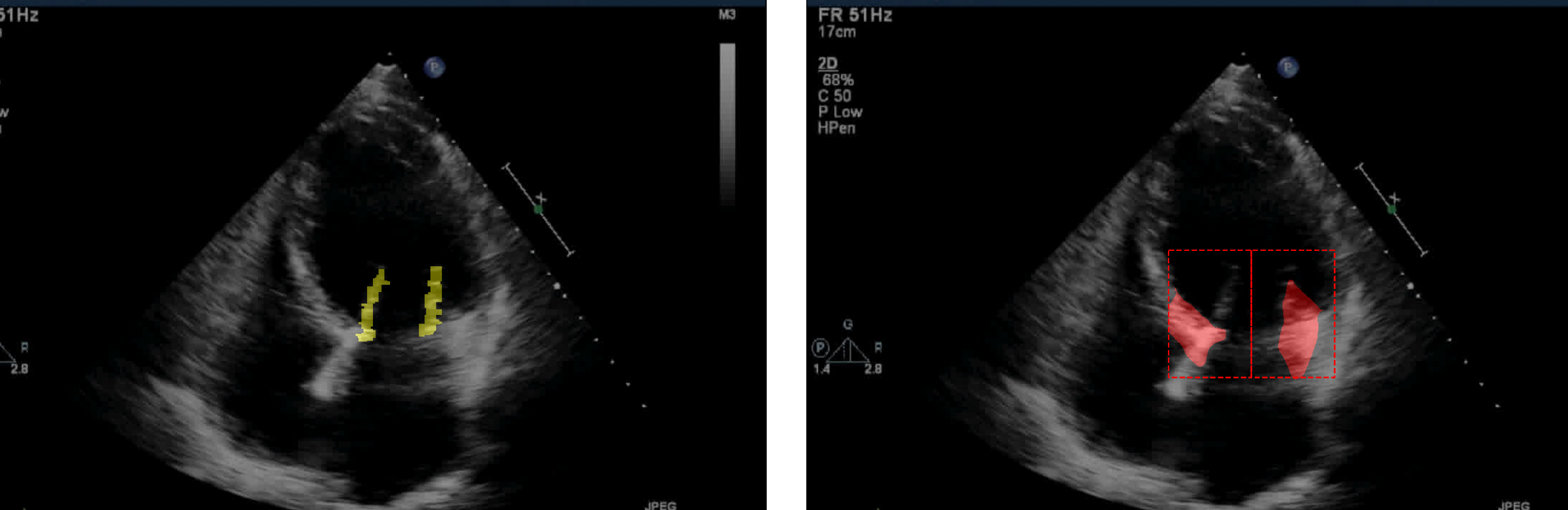} 
		\caption{}
		\label{fig:ac_roi_init}
	\end{subfigure}
	\begin{subfigure}{\textwidth}
		\centering
		\includegraphics[width=0.48\textwidth,height=1.55cm]{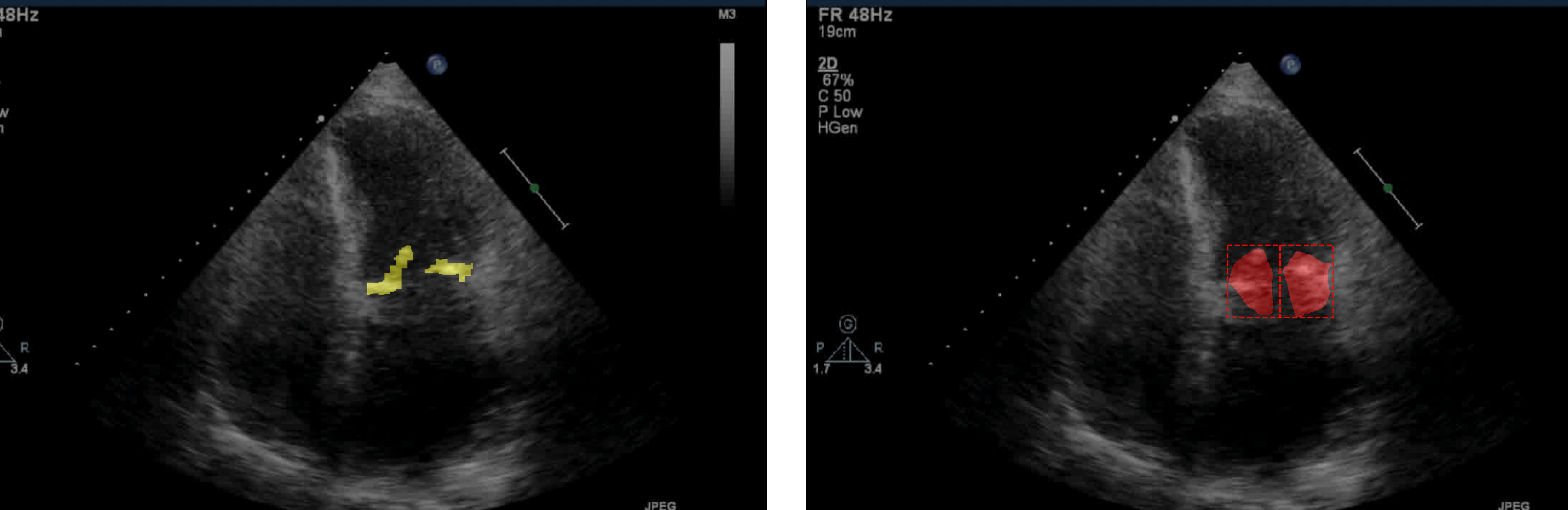}
		\hfill
		\includegraphics[width=0.48\textwidth,height=1.55cm]{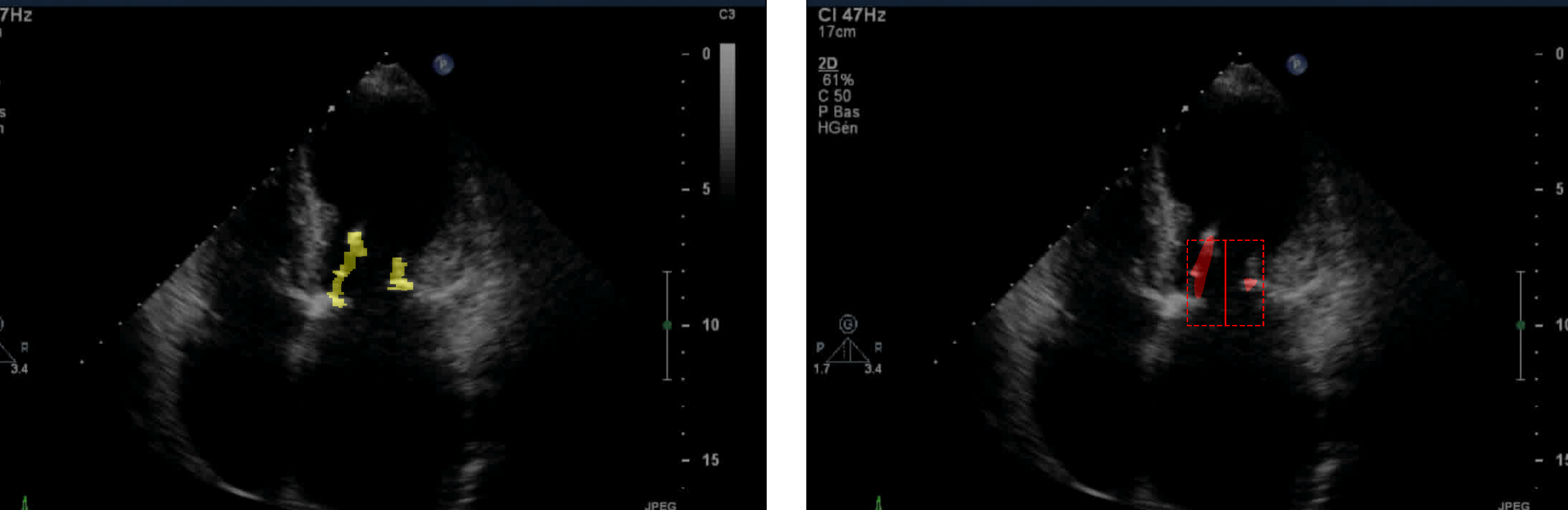}
		\\\vspace{0.1cm}
		\includegraphics[width=0.48\textwidth,height=1.55cm]{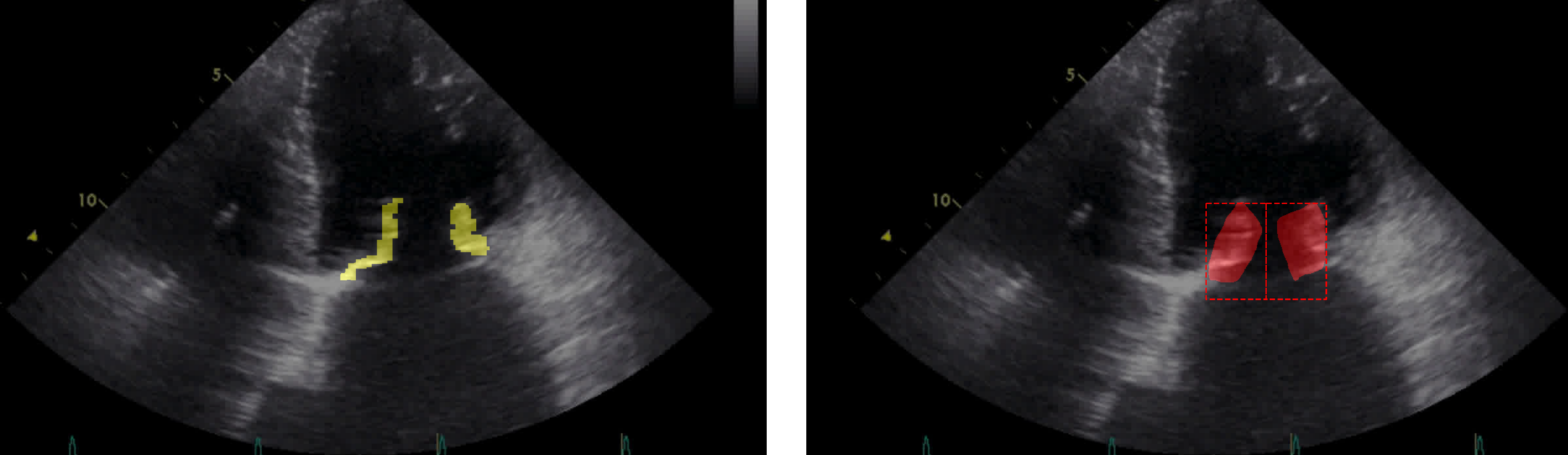}
		\hfill
		\includegraphics[width=0.48\textwidth,height=1.55cm]{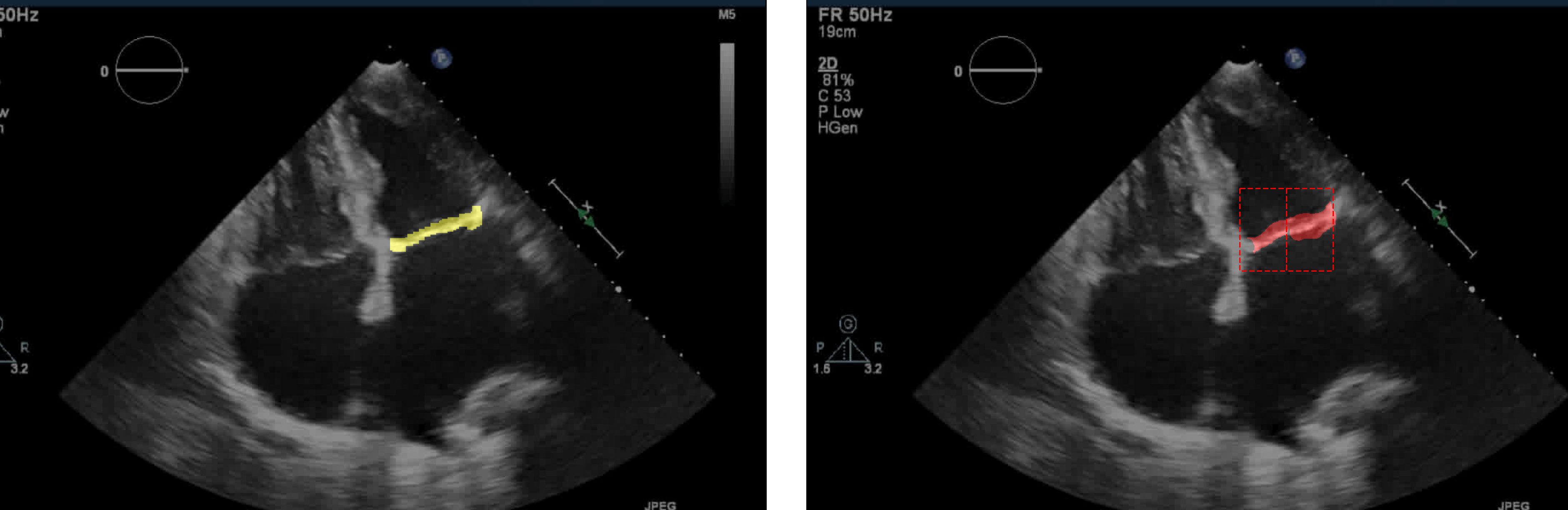}
		\\\vspace{0.1cm}
		\includegraphics[width=0.48\textwidth,height=1.55cm]{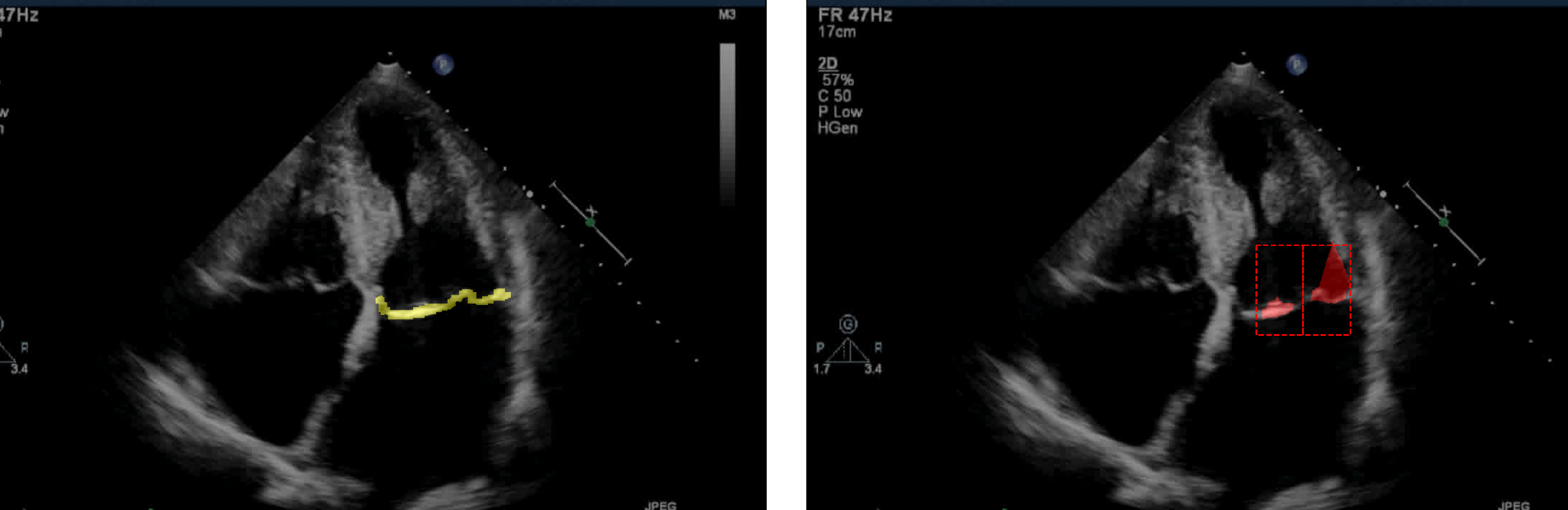}
		\hfill
		\includegraphics[width=0.48\textwidth,height=1.55cm]{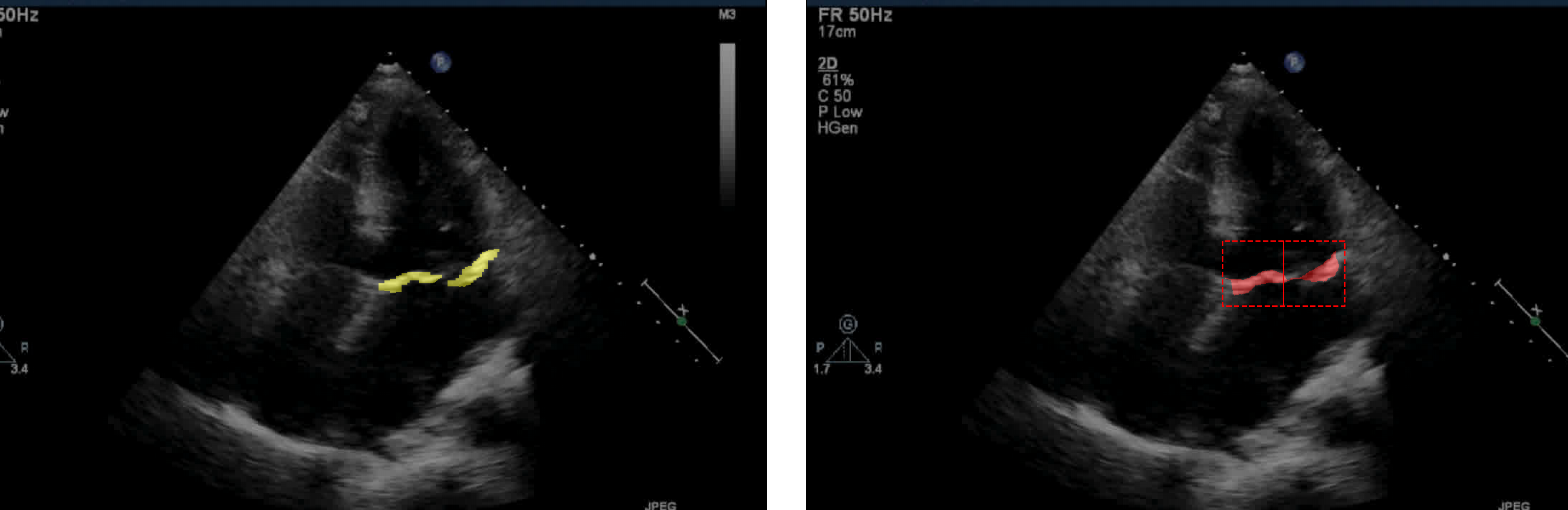}
		\\\vspace{0.1cm}
		\includegraphics[width=0.48\textwidth,height=1.55cm]{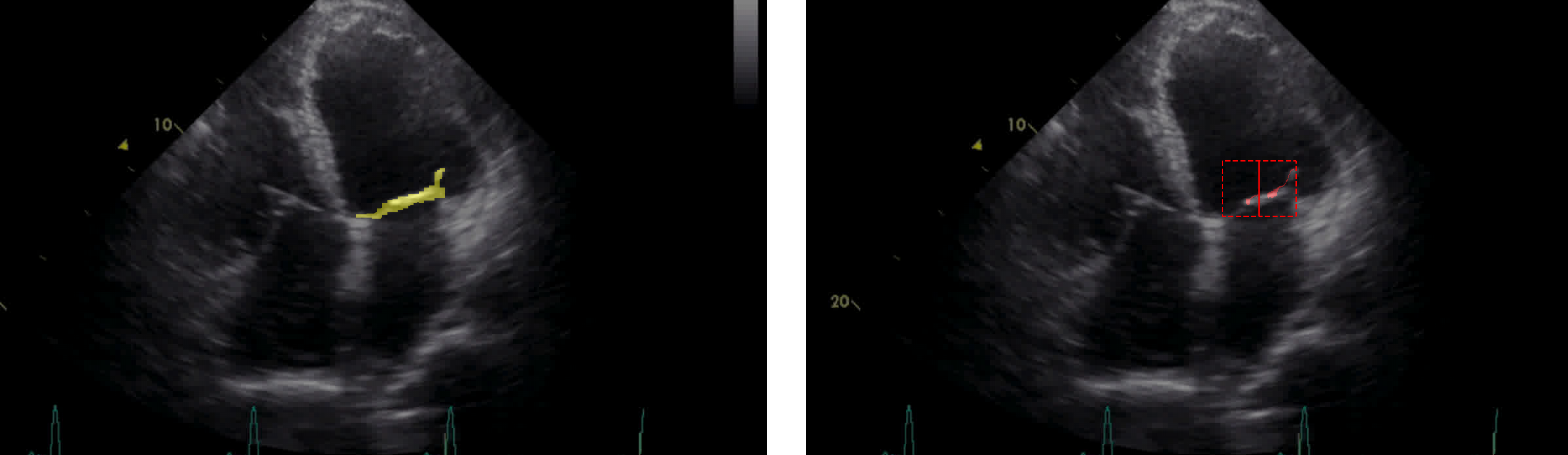}
		\hfill
		\includegraphics[width=0.48\textwidth,height=1.55cm]{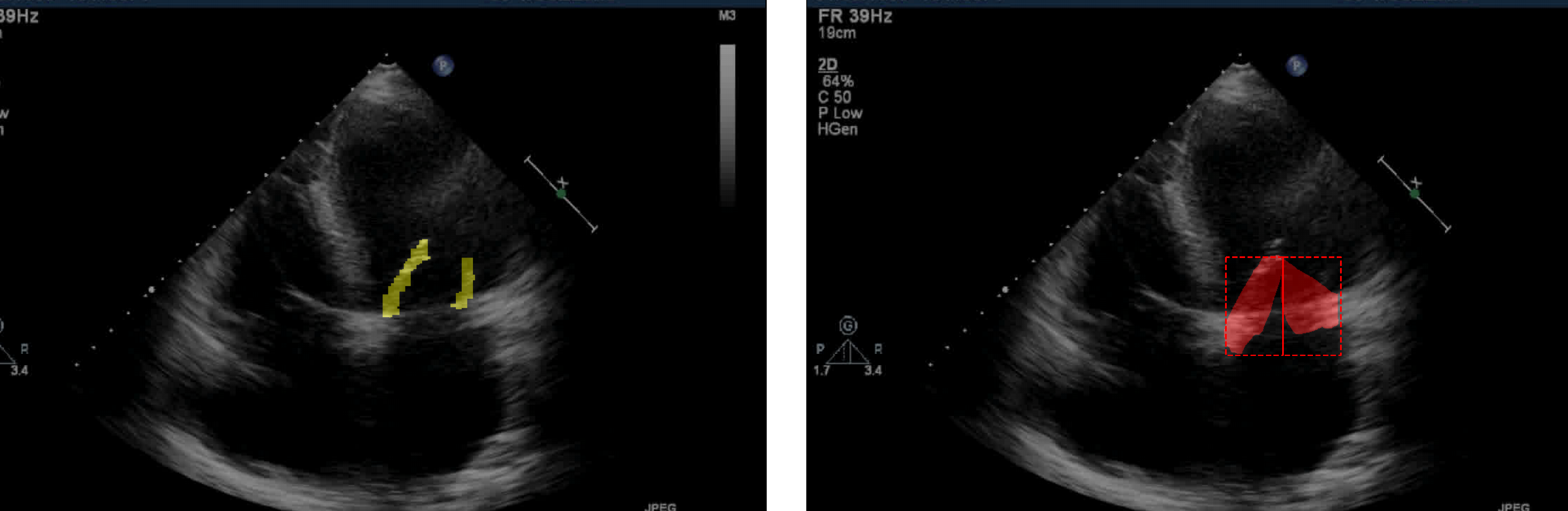}
		\\\vspace{0.1cm}
		\includegraphics[width=0.48\textwidth,height=1.55cm]{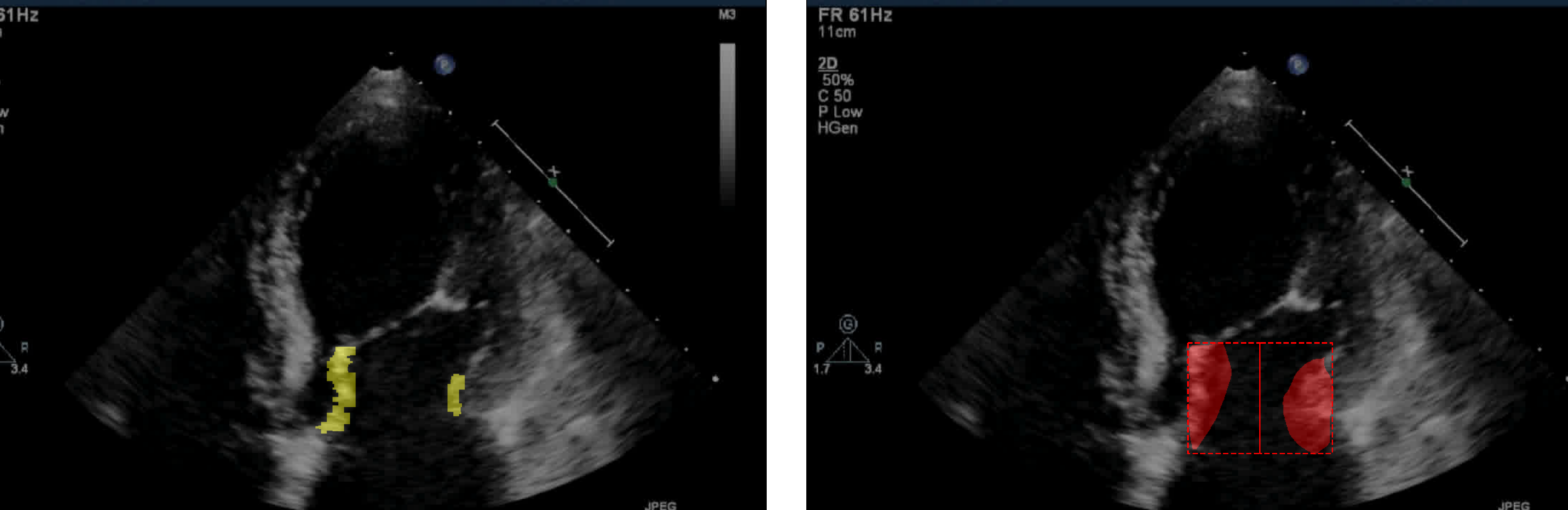}
		\hfill
		\includegraphics[width=0.48\textwidth,height=1.55cm]{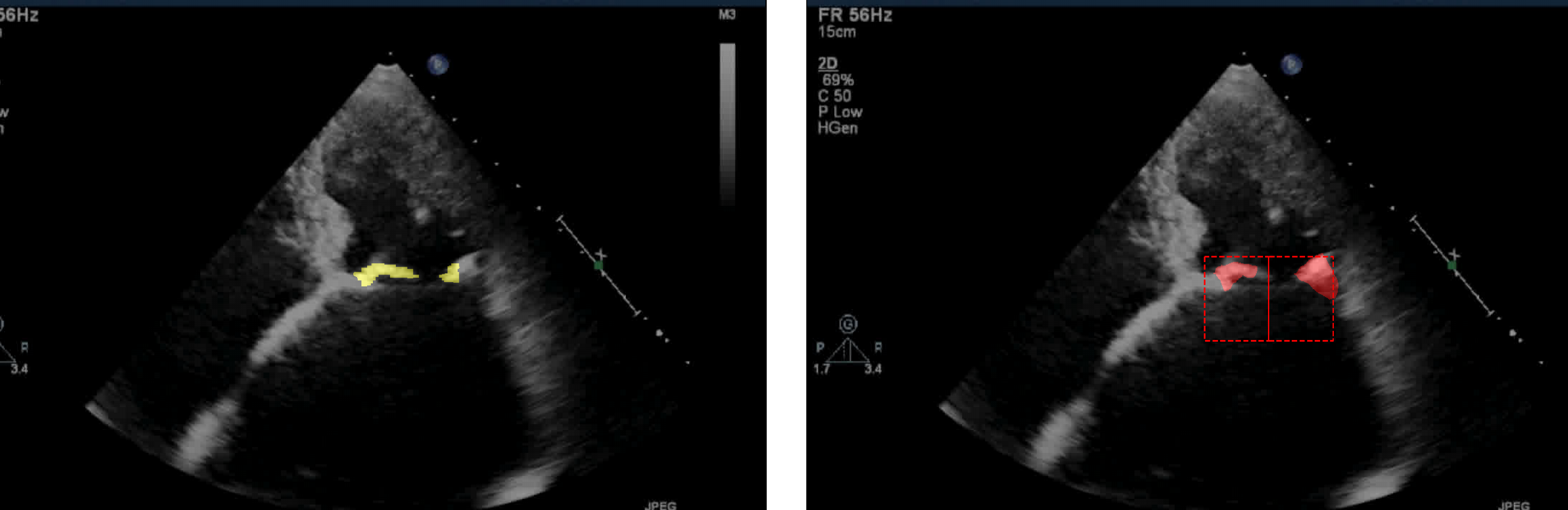}
		\caption{}
		\label{fig:ac_accurate_init}
	\end{subfigure}
	\caption{Examples of the mitral valve segmentation obtained by the active contour method (yellow: ground-truth, red: AC prediction). (a) Initialization given by the ROI provided by medical experts, (b) Improved initialization. Full details in the main text.}
\end{figure}

To have a fair comparison with the other methods, we use as the spline initialization the ROI given for every video. In \Cref{fig:ac_roi_init} we depict a random selection of segmentation predictions and in \Cref{tab:original_cohort_results} and \Cref{tab:echonet_results} (second last row) we report the quantitative performances. It can be observed that the results are particularly poor, with the AC underperforming the unsupervised RNMF benchmark despite using the ROI information. 
A further improvement is given in \Cref{fig:ac_accurate_init} (with quantitative evaluation given again in \Cref{tab:original_cohort_results}, last row), where a more accurate bounding box is computed using prior knowledge of the MV leaflets shapes and sizes. 
The heavy supervision in the form of the initialization improvement gives an important boost to the performance of the AC method (+0.2 in Dice score, \Cref{tab:original_cohort_results}).
Nonetheless, the AC (with naive an accurate initialization) underperforms both unsupervised and more recent supervised methods (based on deep learning), hence can be considered outdated for the task considered here.
\section{Gaussian smoothing loss}
\label{app:gaussian_smoothing}
In order to encode spatio-temporal information in the NeuMF model we here design an additive loss that acts on the the embedding vectors. Let us reshape the collection of embeddings $\{\mathbf{u}^{GMF}_n\}_{n=1}^N$ with $\mathbf{u}^{GMF}_n \in \mathbb{R}_+^{K}$ into a 3D array $\mathbf{U}_{3D}^{GMF} \in \mathbb{R}_+^{w \times h \times K}$ of the same shape of single frame, with an additional dimension (the channel dimension) given by $K$. The same reformatting can be applied also to the other embedding vectors obtaining $\mathbf{U}_{3D}^{MLP} \in \mathbb{R}_+^{w \times h \times K'}$, $\mathbf{V}_{2D}^{GMF} \in \mathbb{R}_+^{T \times K}$ and $\mathbf{V}_{2D}^{MLP} \in \mathbb{R}_+^{T \times K'}$.  Given this reformatting, we can add a loss function that penalizes spatial and temporal variations of the embedding as
\begin{align}
\begin{split}
L_{gs} = \| \mathbf{U}_{3D}^{GMF} - \mathbf{Ker}_{3D} * \mathbf{U}_{3D}^{GMF}  \|_F^2 + \| \mathbf{U}_{3D}^{MLP} - \mathbf{Ker}_{3D} * \mathbf{U}_{3D}^{MLP}  \|_F^2 +  \\
+ \| \mathbf{V}_{2D}^{GMF} - \mathbf{Ker}_{2D} * \mathbf{V}_{2D}^{GMF}  \|_F^2 + \| \mathbf{V}_{2D}^{MLP} - \mathbf{Ker}_{2D} * \mathbf{V}_{2D}^{MLP}  \|_F^2
\end{split}
\end{align}
where $\mathbf{Ker}_{3D}$ and $\mathbf{Ker}_{2D}$ are respectively a 3D kernel (spatial kernel + channel dimension) and a 2D kernel (temporal kernel + channel dimension). 
\begin{figure}[h]
	\centering
	\begin{minipage}{0.9\textwidth}
	\begin{subfigure}[t]{0.62\textwidth}
		\centering
		\includegraphics[width=\textwidth]{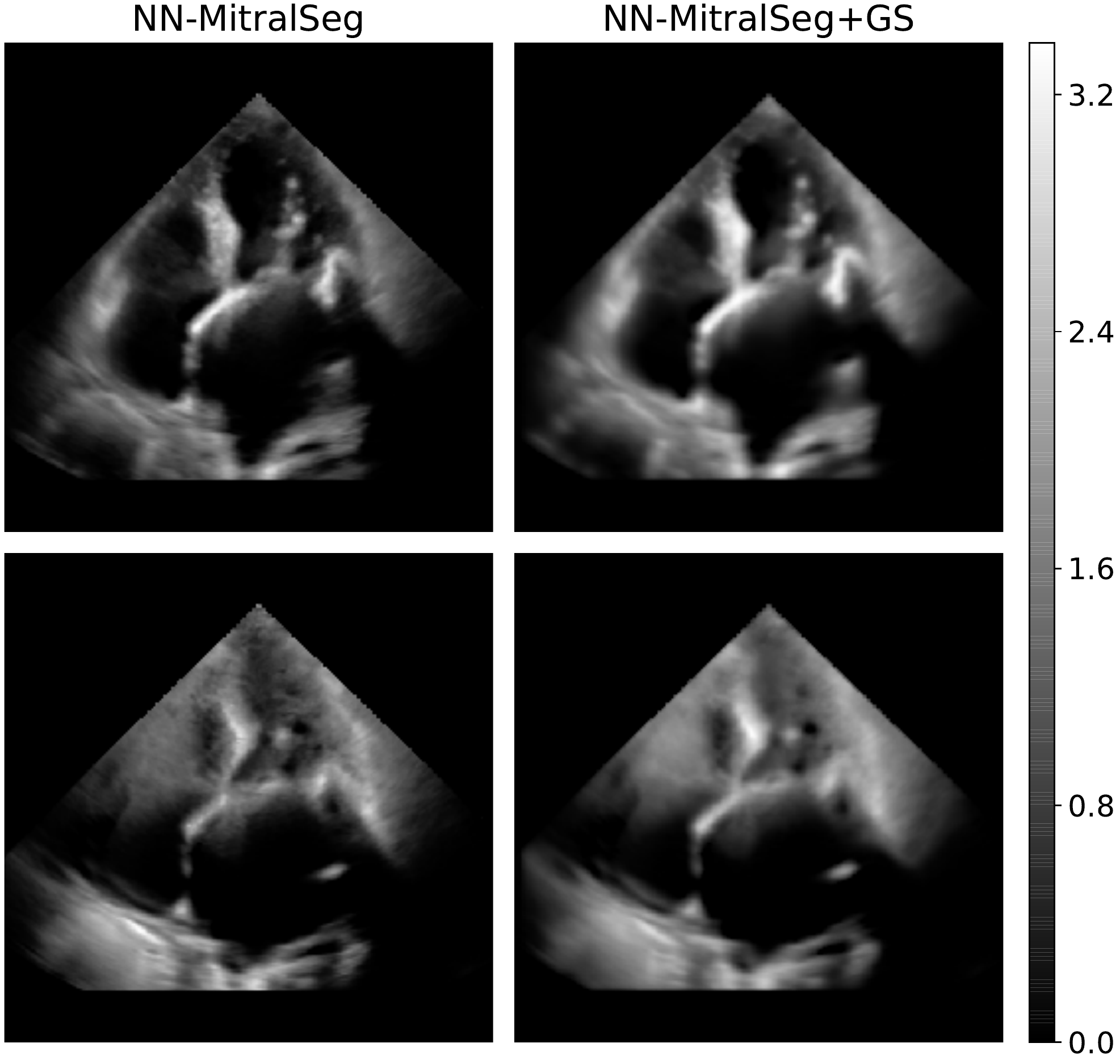}
		\caption{}
		\label{fig:smoothing_embeddings}
	\end{subfigure}
	\hfill
	\begin{subfigure}[t]{0.37\textwidth}
		\centering
		\includegraphics[width=\textwidth]{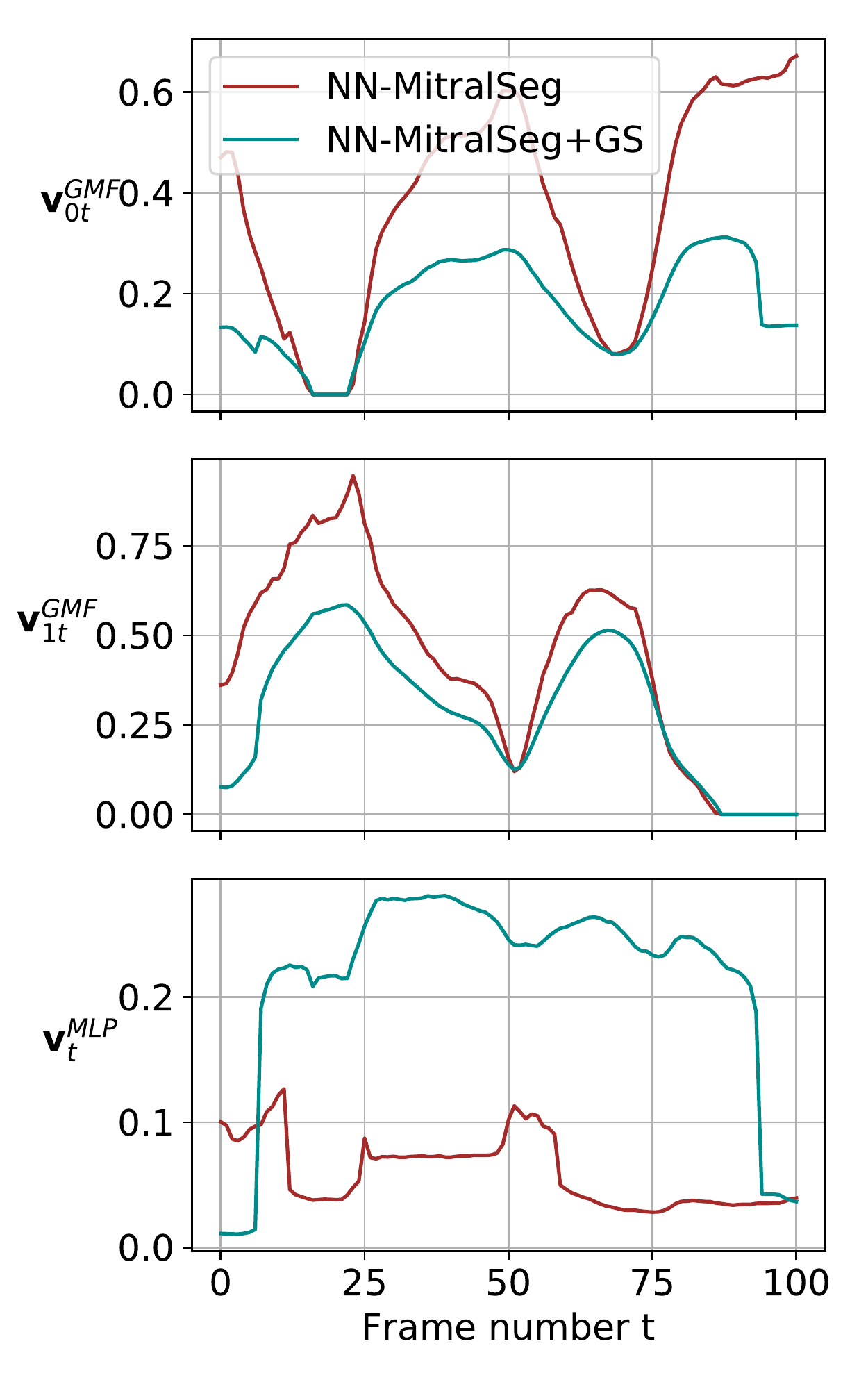}
		\caption{}
		\label{fig:smoothing_time_series}
	\end{subfigure}
	\end{minipage}
	\caption{Comparison of the embedding vectors for the NN-MitralSeg method and the same method with an additional kernel convolutional loss on both the spatial (a) and temporal (b) domain. The embeddings are obtained in both cases at the end of the training.}
	\label{fig:gaussian_smoothing}
\end{figure}
In the experiment, we used Gaussian kernels with a kernel size of 15 and variance fixed to 1, identical for all the channels, and we compare the effect of the additive loss on the NN-MitralSeg method (NeuMF factorization, with temporal masking and optical flow window detection). In \Cref{fig:gaussian_smoothing}, we can observe the qualitative effect of the $L_{gs}$ loss, that has a smoothing effect on both the spatial (\Cref{fig:smoothing_embeddings}) and temporal embeddings (\Cref{fig:smoothing_time_series}). We can further observe in \Cref{fig:smoothing_time_series} how the loss $L_{gs}$ penalizes also the magnitude of the embedding hence acting as an additional $\ell_2$ regularization, with the embeddings $\mathbf{v}^{GMF}$ obtained with Gaussian smoothing loss being consistently smaller than the original ones.
Quantitative results are given in \Cref{tab:original_cohort_results}, where we can observe that the spatio-temporal continuity loss decreases the performance in all metrics, suggesting that the additional constraint impairs the expressiveness of the model.
\end{appendices}

\end{document}